\documentclass[reprint]{aastex62}
\accepted{for publication in The Astrophysical Journal}

\shorttitle{Chakrabarty \& Sengupta: Polarization in Light Reflected from Exoplanets}
\shortauthors{Chakrabarty and Sengupta}
\usepackage{natbib}
\usepackage{amsmath}
\usepackage{wasysym}
\usepackage{epstopdf}

\begin{document}

\renewcommand{\vec}[1]{\boldsymbol{#1}}
\newcommand{\mvec}[1]{$\boldsymbol{#1}$}
\renewcommand{\vector}[4]{\begin{bmatrix} #1\\ #2\\ #3\\ #4\end{bmatrix}}\renewcommand{\matrix}[6]{\begin{bmatrix} #1&#2&0&0\\#2&#3&0&0\\0&0&#4&#5\\0&0&$-$#5&#6 \end{bmatrix}}
\newcommand{\matrixz}[5]{\begin{bmatrix} #1&#2&0&0\\#2&#3&0&0\\0&0&#4&0\\0&0&0&#5 \end{bmatrix}}
\newcommand{\I}{$I$ }
\newcommand{\Q}{$Q$ }
\newcommand{\U}{$U$ }
\newcommand{\V}{$V$ }
\renewcommand{\deg}{$^\circ$ }
\renewcommand{\arccos}[1]{\cos^{-1}\left(#1\right)}
\newcommand{\muphi}{$(\mu,\phi) $}
\newcommand{\muphip}{$(\mu',\phi') $}
\newcommand{\muphiz}{$(-\mu_0,\phi_0) $}
\newcommand{\Mphi}{$(M,\Phi) $}
\newcommand{\cont}{$F_{obs}/F_{*,obs}$}
\newcommand{\alb}{$F/F_0$}
\newcommand{\alba}{$F(\alpha)/F_0$}
\newcommand{\pdisk}{$P_{disk}$}
\newcommand{\pobs}{$P_{obs}$}
 
\title{Generic Models for Disk-Resolved and Disk-Integrated Phase Dependent Linear Polarization of Light Reflected from Exoplanets}

\correspondingauthor{Aritra Chakrabarty}
\email{aritra@iiap.res.in}

\author[0000-0001-6703-0798]{Aritra Chakrabarty}
\affil{Indian Institute of Astrophysics, Koramangala 2nd Block,
Sarjapura Road, Bangalore 560034, India}
\affil{University of Calcutta, Salt Lake City, JD-2 Kolkata 700098, India}

\author[0000-0002-6176-3816]{Sujan Sengupta}
\affil{Indian Institute of Astrophysics, 
Koramangala 2nd Block, Sarjapura Road, Bangalore 560034, India}

\begin{abstract}
Similar to the case of solar system planets, reflected starlight from exoplanets is expected to be polarized due to atmospheric scattering and the net disk integrated polarization should be non-zero owing to the asymmetrical illumination of the planetary disk. The computation of the disk-integrated reflected flux and its state of polarization involves techniques for the calculation of the local reflection matrices as well as the numerical recipes for integration over the planetary disks. In this paper, we present a novel approach to calculate the azimuth-dependent reflected intensity vectors at each location on the planetary disk divided into grids. We achieve this by solving the vector radiative transfer equations that describe linear polarization. Our calculations incorporate self-consistent atmospheric models of exoplanets over a wide range of equilibrium temperature, surface gravity, atmospheric composition, and cloud structure. A comparison of the flux and the amount of polarization calculated by considering both single and multiple scattering exhibits the effect of depolarization due to multiple scattering of light depending on the scattering albedo of the atmosphere. We have benchmarked our basic calculations against some of the existing models. We have also presented our models for the hot Jupiter HD 189733 b, indicating the level of precision required by future observations to detect the polarization of this planet in the optical and near-infrared wavelength region. The generic nature and the accuracy offered by our models make them an effective tool for modeling the future observations of the polarized light reflected from exoplanets.
\end{abstract}

\section{Introduction} \label{sec:intro}

After the discovery of thousands of exoplanets, new direct and indirect techniques are now being sought for the characterization of the atmospheres of these planets. Techniques such as transit spectroscopy \citep[e.g.,][]{chakrabarty20,sengupta20,sing16,waldmann15a,waldmann15b,tinetti07}, eclipse spectroscopy \citep[e.g.,][]{nikolov18,evans17,stevenson17,stevenson14a}, orbital phase curves using infrared photometry \citep[e.g.,][]{demory16,knutson12,knutson07,crossfield12,crossfield10,harrington06}, etc. have conveyed valuable information about the properties of the exoplanetary atmospheres, such as atmospheric compositions, cloud coverage, etc. The polarimetric techniques of studying the atmospheres of the exoplanets are increasingly gaining importance over the past two decades. Long before the exoplanets were discovered, polarimetric studies of the planets began with the observations of the solar system objects, \citep[e.g.,][etc.]{coffeen69, coffeen69a, bugaenko73, hall74, michalski77, west83, smith84, tomasko84, west90} and such investigations are still being continued \citep[e.g.,][etc.]{braak02, gisler03, schmid06, joos07}. Detection of polarization of reflected light from exoplanets is challenging as the signal-to-noise ratio (S/N) in this case is extremely low compared to the case of the solar-system planets. In this regard, theoretical modeling of the polarized reflected light from a planet plays a crucial role in order to guide the design for an appropriate instrument on board the existing and future next generation ground- as well as space-bound telescopes.

Model predictions and recent observations suggest that the light from an exoplanet can be polarized due to various processes observable in different wavelength bands. For example, the thermal emission from a directly imaged giant exoplanet can be polarized in the infrared region \citep[e.g.,][]{jensen20, sanghavi18, stolker17, sengupta13, marley11} due to reasons such as rotation induced oblateness of the planet, nonuniform or banded cloud patterns, gravitational darkening, etc., as is found in the case of the brown dwarfs \citep[e.g.,][]{jensen20,millar20,sanghavi19,sengupta16a,sengupta10,sengupta09}.  On the other hand, time-dependent polarization may arise when a planet transits a dusty ultra-cool dwarf (a red dwarf or a brown dwarf), or when an Exomoon transits a cloudy (or hazy) directly imaged exoplanet \citep[e.g.,][]{milespaez19,sengupta18,sengupta16}.

Our present work focuses on the polarization detectable in the light reflected from an exoplanet. The reflected light gets polarized due to the scattering process in the atmosphere depending on the scattering mechanism and the scatterers. In the context of exoplanets, current technology allows us to study the polarization in the reflected light only for the close-in exoplanets. This makes the detection of polarization in the reflected light even more sensitive ($\sim$ parts per million level), as the net detectable degree of linear polarization for an unresolved planet-star pair is the linearly polarized flux from the planet normalized by both the total reflected flux from the planet and the stellar flux, assuming the radiation of the host star is unpolarized. However, the new generation of high-precision polarimeters \citep[e.g.,][]{hough06, wiktorowicz08, wiktorowicz15, bailey15} are sensitive to polarization at parts per million (ppm).

\cite{berdyugina08} and \cite{berdyugina11} reported the detection of polarization phase curves due to reflection from the planet HD 189733 b in the visible bands with amplitude levels of 100-200 ppm. This was later unconfirmed by further observations \citep[e.g.,][]{lucas09, wiktorowicz15a, bott16}. \cite{sengupta08} also demonstrated that even if we consider only single scattering in a purely scattering atmosphere of that planet, the amplitude of polarization cannot account for the high degree of polarization reported. In fact, the inclusion of multiple scattering in the atmosphere leads to even a lower amount of polarization. Models by \cite{seager00, buenzli09}; etc. calculate the polarized reflected light considering multiple scattering in the atmosphere using a Monte-Carlo based approach. On the other hand, models by \cite{abhyankar70, madhusudhan12, natraj12}; etc. follow the analytical approach assuming a semi-infinite atmosphere that obeys the principle of invariance (see, e.g., \cite{chandrasekhar60}). However, a numerical approach is essential for modeling the polarization profiles as it perfectly trades off between accuracy and computing time, rendering it useful for the retrieval of the atmospheric properties from actual observations. \cite{stam06, karalidi12, karalidi13, rossi18}; etc. followed the adding-doubling method \citep[e.g.,][]{hovenier04} to calculate the local reflection matrices and integrated the polarization over the disk using the numerical recipe suggested by \cite{stam06}. \cite{bailey18, kopparla16}; etc. follow the numerical vector RT model \texttt{VLIDORT} \citep{spurr06} for the prediction of the polarization phase curves of the light reflected from the close-in exoplanets. On the other hand, \cite{sanghavi21} use the conics-based Matrix Operator Method \citep{sanghavi14} to calculate the local Stokes vectors and a numerical technique \citep{sanghavi18} to calculate the disk-integrated polarized reflected light from the wide-orbit self-luminous exoplanets.

In this paper, we present new self-consistent models for the disk-resolved as well as the disk-integrated reflected flux and its state of polarization calculated by using the state-of-the-art numerical techniques developed by us. This is a part of the generic Python-based package that we have recently developed. The main component of the package is the numerical code based on the discrete space theory developed by \cite{peraiah73} that solves the multiple scattering vector radiative transfer equations. The numerical code has been used to solve the scalar and vector radiative transfer equations in order to calculate the polarized spectra of the cloudy brown dwarfs and the self-luminous directly imaged exoplanets \citep{sengupta09, sengupta10, marley11, sengupta16, sengupta16a, sengupta18} and the optical transmission spectra of the transiting exoplanets \citep{sengupta20}. All the Fortran-based code modules have been rewritten and modified in Python before including them in the package. The package is capable of calculating all the atmospheric processes to predict the scalar and vector (polarized) form of the observable spectra, viz. transmission spectra, emission spectra, and reflection spectra, as well as the phase curves. This Python-based package has been applied to model the transmission and the emission spectra of the hot Jupiters presented in \cite{chakrabarty20}. We have also added libraries to the package to calculate the disk-resolved and phase-resolved flux and polarization and included numerical techniques to integrate the flux and polarization over the planetary disk.  

Using this package, we calculate the reflected flux and polarization from the planets with different orbital inclination angles and different orbital phases. We do not expand the phase matrices or the intensity vectors in generalized spherical functions, unlike all other models that use numerical techniques. Instead, we calculate the azimuth-dependent intensity vectors by solving the vector radiative transfer equations and then integrate over the disk to get the disk-integrated vector flux. For the assessment of these models, we have benchmarked some of our results against previously published results. Our model works over a broad range of physical properties of the planets such as equilibrium temperature, surface gravity, orbital inclination angle, planetary radius, etc. as well as cloud structures and chemical compositions that can describe the atmospheres of reflecting exoplanets. For the transiting planets i.e. planets with orbital inclination angle $\sim90^\circ$, these physical properties can be precisely estimated from transit photometric observations \citep[e.g.,][]{chakrabarty19, addison19, saha21}. On the other hand, the course knowledge of the atmospheric chemical composition and cloud structure estimated from observation and modeling of transmission spectra \citep[e.g.,][]{wilson21, chakrabarty20, sengupta20, sing16} or emission spectra \citep[e.g.,][]{nikolov18, evans17} of the transiting exoplanets helps us develop models for the polarized reflected light observable from those planets. Here, we consider both single scattering and multiple scattering in the atmosphere due to constituent gas molecules and cloud particles and present the effect of depolarization due to multiple scattering in the atmosphere. 

Spectroscopic observations \citep[e.g.,][]{redfield08, sing16, turner20, chen20} suggest that the atmospheres of the hot Jupiters can contain a significant amount of the atomic absorbers such as Na, K, Ca, etc. Hence, we have studied the effect of these absorbers on the albedo and the degree of polarization of the hot Jupiters. We also present the model predictions for the albedo and the polarization phase curves of the hot Jupiter HD 189733 b.

Section~\ref{sec:ivpm} discusses the basic concepts regarding the intensity vectors and the scattering phase matrices used in our calculations. Section~\ref{sec:inc-alpha} describes the dependence of the planetary phase on its orbital inclination angle and orbital phase angle. We elaborate the technique of solving the vector radiative transfer equations in Section~\ref{sec:vrt}. In Section~\ref{sec:diskint}, we describe the numerical recipe for integrating the intensity vectors over the planetary disk. Section~\ref{sec:benchmark} discusses the results we have obtained from benchmarking our calculations against previously published results. Section~\ref{sec:atmosphere} describes the atmospheric models of the exoplanets we have adopted. In Section~\ref{sec:effects-multi}, we discuss the effects due to multiple scattering of the internal radiations on the calculated flux and polarization. We present our models developed for the exoplanet HD 189733 b in Section~\ref{sec:hd189733b}. In Section~\ref{sec:rd}, we discuss and interpret the results obtained and conclude the significant points in Section~\ref{sec:con}.

\section{Intensity Vector and Scattering Phase Matrix} \label{sec:ivpm}

\begin{figure}[!ht]
\centering
\includegraphics[scale=0.48,angle=0]{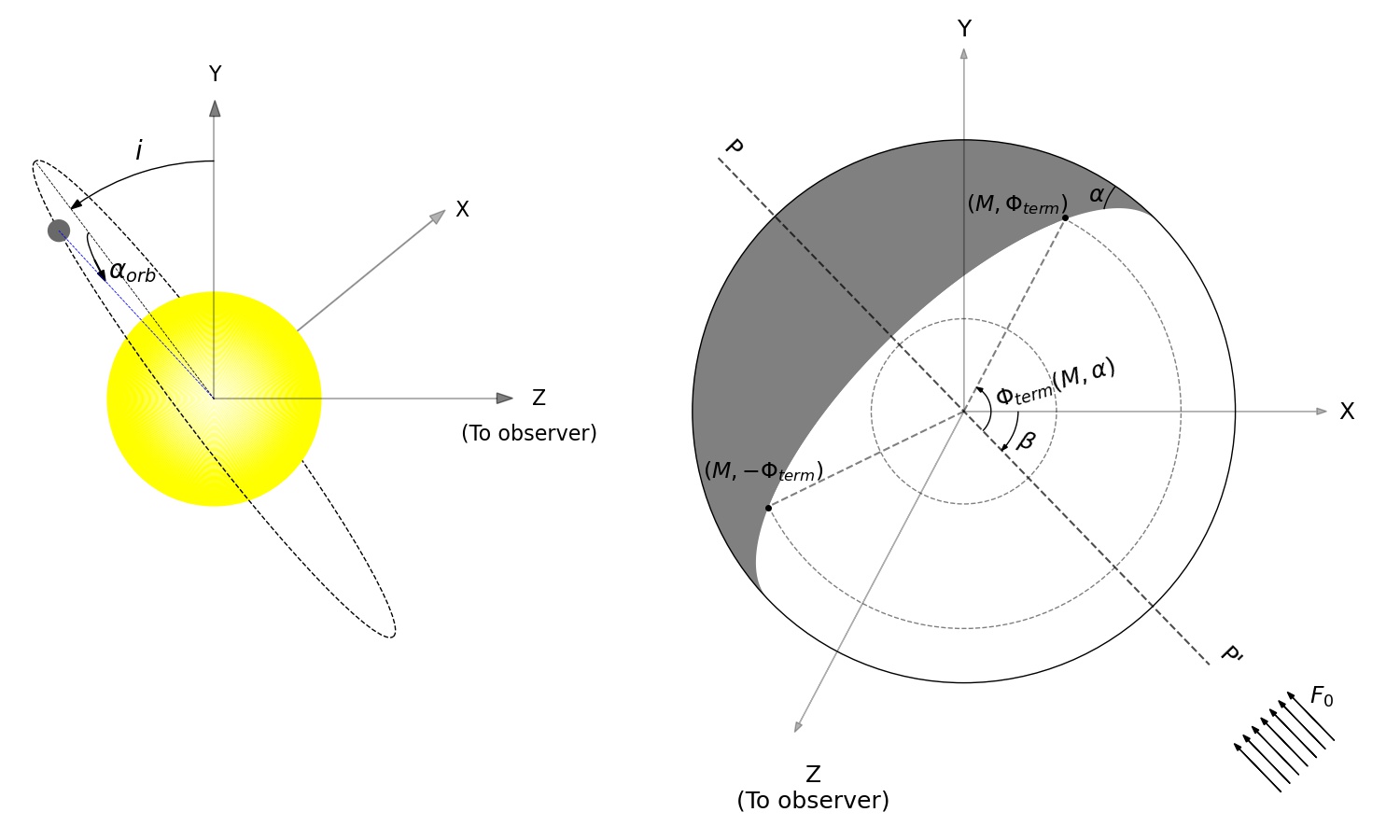}
\caption{Left: The side-view of the orbital position of a planet for an inclination angle, $i=30$\deg\!\! with respect to the XY-plane and a particular orbital phase, $\alpha_{orb}=40$\deg\!\! as seen by the observer (along Z-axis). 
Right: The detailed front view of the dayside and the night-side of a planet as seen by the observer (along Z-axis) for the same inclination angle and orbital phase. The effective phase angle ($\alpha$) of the planet as seen by the observer is a function of $i$ and $\alpha_{orb}$ (see Equation~\ref{eq:alpha}). The direction of the incident stellar flux $F_0$ is parallel to the planetary scattering plane PP$^\prime$ which is at an angle $\beta$ with respect to the XZ-plane. Any point on the terminator is represented with the cosine of the polar angle ($M=\cos(N)$) and the azimuthal angle $\Phi_{term}(M,\alpha)$ (see Equation~\ref{eq:phiterm}).
\label{fig:daynight-detail}}
\end{figure}

The wavelength dependent intensity vector or Stokes vector \mvec{I} is expressed as \citep[e.g.,][]{chandrasekhar60, hovenier04, sengupta06},
\begin{equation} \label{eq:ivec}
\vec{I} = \vector{I}{Q}{U}{V},
\end{equation}

where \I denotes the total intensity, \Q, \U, and \V denote the polarized intensities \citep{chandrasekhar60}. For linear polarization \V= 0. 
\Q is taken to be positive along the $XZ$-plane and negative along the $YZ$-plane (see figure~\ref{fig:daynight-detail}), Z being the direction of propagation of light.
On the other hand, \U is taken to be positive and negative along the planes making +45\deg (counter-clockwise) and -45\deg (clockwise) respectively with the $XZ$ plane.  

When starlight is incident on a planet's atmosphere, a fraction of the light gets scattered off in the atmosphere in different directions depending on the single scattering albedo ($\omega$) of the atmosphere and the scattering mechanism, which in turn depends on the chemical species in the atmosphere that scatters the light. For planets with a thick atmosphere, the amount of reflected light and its polarization solely depend on this scattering process as the surface reflectivity can be ignored in such cases. The quantities \Q and \U and hence, \mvec{I} are defined with respect to a plane of reference. As the plane of reference is rotated by an angle $\beta$ (counter-clockwise), the new intensity vector is obtained by multiplying the original intensity vector with a rotation matrix \mvec{L}, given by,
\begin{equation}
\vec{L}(\beta) = {\begin{bmatrix} 1&0&0&0\\0&\cos 2\beta&\sin 2\beta&0\\0&-\sin 2\beta&\cos 2\beta&0\\0&0&0&1 \end{bmatrix}}
\end{equation}

If an intensity vector \mvec{I} gets scattered to an angle $\Theta$ from its initial direction then the scattered intensity vector in the new direction defined with respect to the same reference plane can be expressed as,
\begin{equation} \label{eq:phmat}
\vec{I}(\Theta) = \vec{P}(\Theta)\vec{I},
\end{equation}

where $\vec{P(\Theta})$ denotes the phase matrix of the scattering process which can be broadly expressed as \citep[e.g.,][]{hovenier69, hovenier70, hansen74, stam06}
\begin{equation} \label{eq:phmat-gen}
\vec{P}(\Theta) = \matrix{P_{11}(\Theta)}{P_{12}(\Theta)}{P_{22}(\Theta)}{P_{33}(\Theta)}{P_{34}(\Theta)}{P_{44}(\Theta)}.
\end{equation}

Here, $P_{11}(\Theta)$ is known as the phase function and $P_{12}(\Theta)/P_{11}(\Theta)$ is known as the single scattering polarization \citep[e.g.,][]{stam06}. For a cloud-free atmosphere, the gas molecules scatter the light and the corresponding phase matrix is given by the Rayleigh phase matrix. For an isotropic Rayleigh scattering, this phase matrix can be expressed as,
\begin{equation} \label{eq:phmat-ray}
\vec{P_{mol}}(\Theta) = \frac{3}{4}\matrixz{(1+\cos^2\Theta)}{$-$\sin^2\Theta}{(1+\cos^2\Theta)}{2\cos\Theta}{2\cos\Theta}
\end{equation}

However, to express the scattering phase matrix for anisotropic Rayleigh particles in random orientations, a depolarization factor needs to be introduced \citep[e.g.,][]{hansen74, hovenier04}. The phase matrix corresponding to the scattering by the cloud particles is represented by Mie phase matrix, expressed as,
\begin{equation} \label{eq:phmat-mie}
\vec{P_{cld}}(\Theta) = \frac{1}{2}\matrix{(S_1S_1^*+S_2S_2^*)}{(S_1S_1^*-S_2S_2^*)}{(S_1S_1^*+S_2S_2^*)}{(S_1S_2^*+S_2S_1^*)}{i(S_1S_2^*-S_2S_1^*)}{(S_1S_2^*+S_2S_1^*)}
\end{equation}
The details of this phase matrix and the terms $S_1$ and $S_2$ can be found in \cite{vandehulst57, hansen74, fowler83}; etc. In the case of scattering by both the gas molecules and the cloud particles, the combined phase matrix can be obtained by adding $\vec{P_{mol}}(\Theta)$ and $\vec{P_{cld}}(\Theta)$ in the ratio of their respective single-scattering albedos, $\omega_{mol}$ and $\omega_{cld}$ respectively.

\section{Inclination Angle, Orbital phase and Phase angle of Planets} \label{sec:inc-alpha}

The reflected planetary radiation and the state of polarization are functions of the phase angle ($\alpha$) of the planet \citep{stam06, madhusudhan12, kopparla16, batalha19, sanghavi21} which in turn is a function of both the orbital inclination angle ($i$) and the orbital phase angle ($\alpha_{orb}$) (see Figure~\ref{fig:daynight-detail}). Here, the orbital phase angle is defined in terms of the angular displacement from the planet's orbital position at the secondary eclipse (or, the superior conjunction, in the case of the solar-system planets). Usually, the orbits of the close-in exoplanets are circular and so the time of periastron is the same as the time of the secondary eclipse (or, superior conjunction). Hence, the orbital phase, the true anomaly, and the mean anomaly are all the same in this case \citep[e.g.,][]{madhusudhan12}. Under such a situation, the phase angle of the planet as seen by the observer can be expressed as,
\begin{equation} \label{eq:alpha}
\cos \alpha = \sin i \cos \alpha_{orb}
\end{equation}
For $i=90^\circ$, i.e. when a planet is viewed edge-on, the phase angle and orbital phase become equal. For $i=0^\circ$, i.e. when the planet is viewed face-on, the orbital phase is defined to be zero when the planetary scattering plane makes an angle of exactly +90$^\circ$ with respect to the optical plane of observation (i.e., along the XZ-plane).

\section{Numerical Solution of the Vector Radiative Transfer Equations} \label{sec:vrt}

In order to calculate the local radiation field and the local polarization, we solve the vector radiative transfer equations for a vertically inhomogeneous medium. Our numerical code allows the calculations for horizontally inhomogeneous planetary atmospheres. However, we present our models here for a simpler case where we assume a stratified horizontally homogeneous atmosphere. We divide the medium at each location on the disk into a finite number of plane-parallel layers. We then add the radiation field at each layer from bottom to top by using the ``star algorithm" as described in \cite{peraiah73}. The emergent flux and the polarization at all the local points are then integrated over the entire planetary disk. The planetary disk is divided into a grid of polar angles, $N=\arccos{M}$ and azimuthal angles, $\Phi$, considering the disk center at the pole. So, a differential area on the disk centered at ($M,~\Phi$) makes a solid angle, $d\Omega$ with respect to the observer and is given by,
\begin{equation} \label{eq:domega}
d\Omega = \frac{R_P^2}{D^2}M dM d\Phi,
\end{equation}

where $R_P$ is the radius of the planet and $D$ is the distance between the planetary system and the observer ($D>>R_P)$.

At each point on the grid, the directions of propagation of light are again divided into a grid of polar angles, $\nu=\arccos{\mu}$ and azimuthal angles, $\phi$, with respect to the local zenith at that point. We consider $\mu$ to be positive in the outward direction. At each differential area on the disk, we calculate the intensity vectors considering locally plane-parallel layers of the atmosphere. The reflected intensity vectors can be computed by solving the vector radiative transfer (VRT) equations by invoking either single scattering or multiple scatterings in the atmosphere. While single scattering is a good approximation for a thin atmosphere, in most of the exoplanetary atmospheres, multiple scattering is important. The details of both cases are explained in the following subsections. In all of our calculations, we have assumed the planets to be spherical with negligible oblateness and the light from the host stars to be unpolarized.

\subsection{Case I: Single Scattering} \label{sec:vrt/single}

We first calculate the reflected flux and polarization considering only one scattering event at each layer in the atmosphere. The starlight incident on the top of the atmosphere (TOA) of the planet gets transmitted inward through each layer of the atmosphere with exponential decay and we consider only the beams scattered along the direction towards the observer at each layer. The angle between these two beams, i.e., the scattering angle, is constant over the illuminated part of the disk, which is denoted by $\Theta$. Again, along the direction towards the observer at any point ($M,\Phi$) on the illuminated disk, we can set $\mu=M$. Hence, the radiation towards the observer is uniform over $\Phi$ within the illuminated part of the disk. We denote this outgoing intensity vector at any optical depth $\tau$ from the top of atmosphere (TOA) as \mvec{I}($\tau,M,\Theta$). The vector radiative transfer (VRT) equations for this outgoing intensity vector at optical depth $\tau$ from TOA considering only single scattering in the atmosphere can be expressed as,
\begin{equation} \label{eq:vrt-single}
M\frac{d\vec{I}(\tau,M,\Theta)}{d\tau} = \vec{I}(\tau,M,\Theta) - \frac{\omega(\tau)}{4\pi}F_0e^{-\tau/\mu_0}\vec{P}(\Theta),
\end{equation}

where $\omega(\tau)$ is the single scattering albedo of the atmosphere at an optical depth $\tau$ from TOA. \mvec{P}($\Theta$) can be directly used as the phase matrix in Equation~\ref{eq:vrt-single} as long as the incident starlight is unpolarized or the state of polarization is defined with respect to the planetary scattering plane. If the starlight is polarized and defined with respect to any other plane making an angle $\theta$ (counter-clockwise) from the planetary scattering plane, then the phase matrix will be $\vec{P}(\Theta) \vec{L}(\theta)$. However, since we assume the starlight to be unpolarized, we ignore the term $L(\theta)$. The scattered light towards the observer \mvec{I}($\tau,M,\Theta$), however, is defined with respect to the planetary scattering plane regardless.

Clearly, for a phase angle of $\alpha$, the reflected intensity vector from the illuminated disk as a function of \Mphi~ towards the direction of the observer with respect to the planetary scattering plane can be calculated as,
\begin{equation} \label{eq:iscat-single}
\vec{I_{scat}}(M,\Phi) = I(\tau\!=\!0,M,\Theta\!=\!\pi-\alpha). 
\end{equation}

It is worth mentioning here that, although we have considered one scattering event at each layer that amounts to a multiple scattering event at the whole atmosphere, we have fixed the direction of the scattered photon in the same direction, i.e., along the direction of the observer. Hence, effectively the photon suffers single scattering throughout the atmosphere. 

\subsection{Case II: Multiple Scattering} \label{sec:vrt/multi}

In the case of multiple scattering in the atmosphere, we need to consider both single scattering of the incident starlight towards the direction of the observer and multiple scattering of the internal radiations \citep{chandrasekhar60}. The incorporation of the multiple scattering phase matrix in the VRT equations is not as straightforward as in the case of single scattering. An exact solution of the VRT equations involving both single scattering source function and multiple scattering source function calls for an efficient numerical recipe.

For multiple scattering, we consider the intensity vectors as functions of both the location on the disk and the direction of propagation. We denote the intensity vector at a location ($M,\Phi$) on the disk at an optical depth of $\tau$ calculated from TOA towards the direction ($\mu,\phi$) as \mvec{I}($\tau,M,\Phi,\mu,\phi$). However, we need to define a reference plane for \mvec{I}($\tau,M,\Phi,\mu,\phi$). 

The multiple scattering source function in a particular direction \muphi~ is calculated by integrating the scattered light in that direction from all directions ($\mu',\phi'$), where, $-1\leq\mu'\leq1$ and $0\leq\phi'\leq2\pi$. However, the integration cannot be performed unless all the intensity vectors scattered towards ($\mu,\phi$) from ($\mu',\phi'$)  are defined with respect to a common plane of reference. For that reason, we cannot use the phase matrices mentioned in Equations~\ref{eq:phmat}-\ref{eq:phmat-mie} directly, as they are applicable when both the incident beams and the scattered beams are defined with respect to the individual planes of scattering (planes joining the line along ($\mu,\phi$) and each of the lines along ($\mu',\phi'$)). So, following \cite{chandrasekhar60}, we define the incident intensity vectors \mvec{I}($\tau,M,\Phi,\mu',\phi'$) with respect to the local meridian planes along ($\mu',\phi'$) and then change the reference planes from the local meridian planes to the individual planes of scattering (planes joining the line along ($\mu,\phi$) and each of the lines along ($\mu',\phi'$)) to calculate the scattered radiation in the direction of ($\mu,\phi$). Finally, we change the reference planes of the scattered intensity vectors from the individual planes of scattering to the local meridian plane along $(\mu,\phi)$ before integrating them. The entire adjustment can be done on the phase matrix as \citep{chandrasekhar60},

\begin{equation} \label{eq:phmat-vector}
\vec{P_{mer}}(\Theta) = \vec{L}(-(\pi-i_2)) \vec{P}(\Theta) \vec{L}(i_1),
\end{equation}

where $i_1$ denotes the angles between the local meridian planes along \muphip ~and the planes of scattering, and $i_2$ denotes the angles between the planes of scattering and the local meridian plane along \muphi. The values of $i_1$ and $i_2$ can be calculated using the laws of spherical trigonometry (see Fig. 8 of \cite{chandrasekhar60}). The signs in Equation~\ref{eq:phmat-vector} can be found to be opposite to those given in Equation~213 of \cite{chandrasekhar60} because of the fact that we have considered the angle in the counter-clockwise direction to be positive in contrast to the sign convention of \cite{chandrasekhar60}. 

In the case of multiple scattering, instead of denoting the phase matrix as \mvec{P_{mer}}($\Theta$), we denote it as \mvec{P_{mer}}($\mu,\phi;\mu',\phi'$) by replacing the term $\cos\Theta$ in the expression of \mvec{P_{mer}}($\Theta$) with,

\begin{equation} \label{eq:scatangle2d}
\cos\Theta = \mu\mu' + \sqrt{(1-\mu^2)(1-\mu'^2)}\cos~(\phi'-\phi),
\end{equation}

Similarly, considering the incident beam to be directed along \muphiz, the single scattering phase matrix \mvec{P_{mer}}($\mu,\phi;-\mu_0,\phi_0$) can be expressed as,
\begin{equation}
\vec{P_{mer}}(\mu,\phi;-\mu_0,\phi_0) = \vec{L}(-(\pi-i_2)) \vec{P}(\mu,\phi;-\mu_0,\phi_0) \vec{L}(i_1).
\end{equation}

The post-multiplying factor $L(i_1)$  can be ignored if the incident starlight is unpolarized. However, the pre-multiplying factor $L(-(\pi-i_2))$ has to be incorporated to change the reference plane from the plane of scattering to the local meridian plane along \muphi, as the single scattering source function gets added to the multiple scattering source function.

The complete VRT equation can now be expressed as,

\begin{equation} \label{eq:vrt-multi}
\begin{aligned} 
\mu\frac{d\vec{I}(\tau,M,\Phi,\mu,\phi)}{d\tau} = \vec{I}(\tau,M,\Phi,\mu,\phi)
&-\frac{\omega(\tau)}{4\pi}\int_0^{2\pi}\int_{-1}^1 \vec{P_{mer}}(\mu,\phi;\mu',\phi')\vec{I}(\tau,M,\Phi,\mu',\phi') ~d\mu' d\phi'\\
&- \frac{\omega(\tau)}{4\pi}F_0e^{-\tau/\mu_0}\vec{P_{mer}}(\mu,\phi;-\mu_0,\phi_0).
\end{aligned} 
\end{equation}

In order to calculate the reflected light from the above equations at any point \Mphi ~on the illuminated part of the disk towards the observer, we set $\mu=\mu_{obs}$ and $\phi=\phi_{obs}$. Also, at each location \Mphi ~on the disk, we set $\phi_0$ as the zero point for $\phi$. Accordingly, the values of $\mu_0$, $\mu_{obs}$, and $\phi_{obs}$, which are functions of $M,\Phi$ and the phase angle $\alpha$, are given by,
\begin{equation}
\begin{aligned}
\phi_0 &= 0,\\ 
\mu_0 & = M \cos\alpha + \sqrt{1-M^2} \sin\alpha \cos\Phi,\\
\mu_{obs} & = M,\\
\phi_{obs} & = \arccos{\frac{\mu\mu_0-\cos \alpha}{\sqrt{(1-\mu^2)(1-\mu_0^2)}}}.
\end{aligned}
\end{equation}

Also, we need to calculate the reflected intensity vectors with respect to the planetary scattering plane. The angle between the planetary scattering plane and the local meridian plane along the reflected light at each \Mphi ~on the disk is equal to $\Phi$. Hence, the reflected intensity with respect to the planetary scattering plane can be calculated as,
\begin{equation} \label{iscat-multi}
\vec{I_{scat}}(M,\Phi) = \vec{L}(\Phi)\vec{I}(\tau\!=\!0,M,\Phi,\mu\!=\!\mu_{obs},\phi\!=\!\phi_{obs}).
\end{equation}

The intensity vector \mvec{I_{scat}} can be expressed in terms of its component parameters as,

\begin{equation} \label{eq:iscat}
\mathbf{\vec{I_{scat}} = \vector{I_{scat}}{Q_{scat}}{U_{scat}}{V_{scat}},}
\end{equation}

We solve Equation~\ref{eq:vrt-single} and Equation~\ref{eq:vrt-multi} using our numerical techniques based on the discrete space theory \citep{peraiah73}. This technique has been elaborately described in \cite[][etc.]{peraiah73, sengupta09, sengupta20}. This technique has an advantage of getting exact solutions to both the scalar and the vector radiative transfer equations for any direction of incident light and any direction of outgoing light.

\subsection{Transformation of Reference Plane to the Optical Plane of Observation} \label{sec:vrt/obs}

Once the reflected intensity vector \mvec{I_{scat}}($M,\Phi$) is calculated with respect to the planetary scattering plane in a single scattering atmosphere (see Equations~\ref{eq:vrt-single}-\ref{eq:iscat-single}) or in a multiple scattering atmosphere (see Equation~\ref{eq:vrt-multi}-\ref{iscat-multi}), the reference plane needs to be transformed to the optical plane of observation i.e. the optical plane of the Earth-bound polarimeter \citep{stam06}. 

\begin{figure}[!ht]
\centering
\includegraphics[scale=0.5,angle=0]{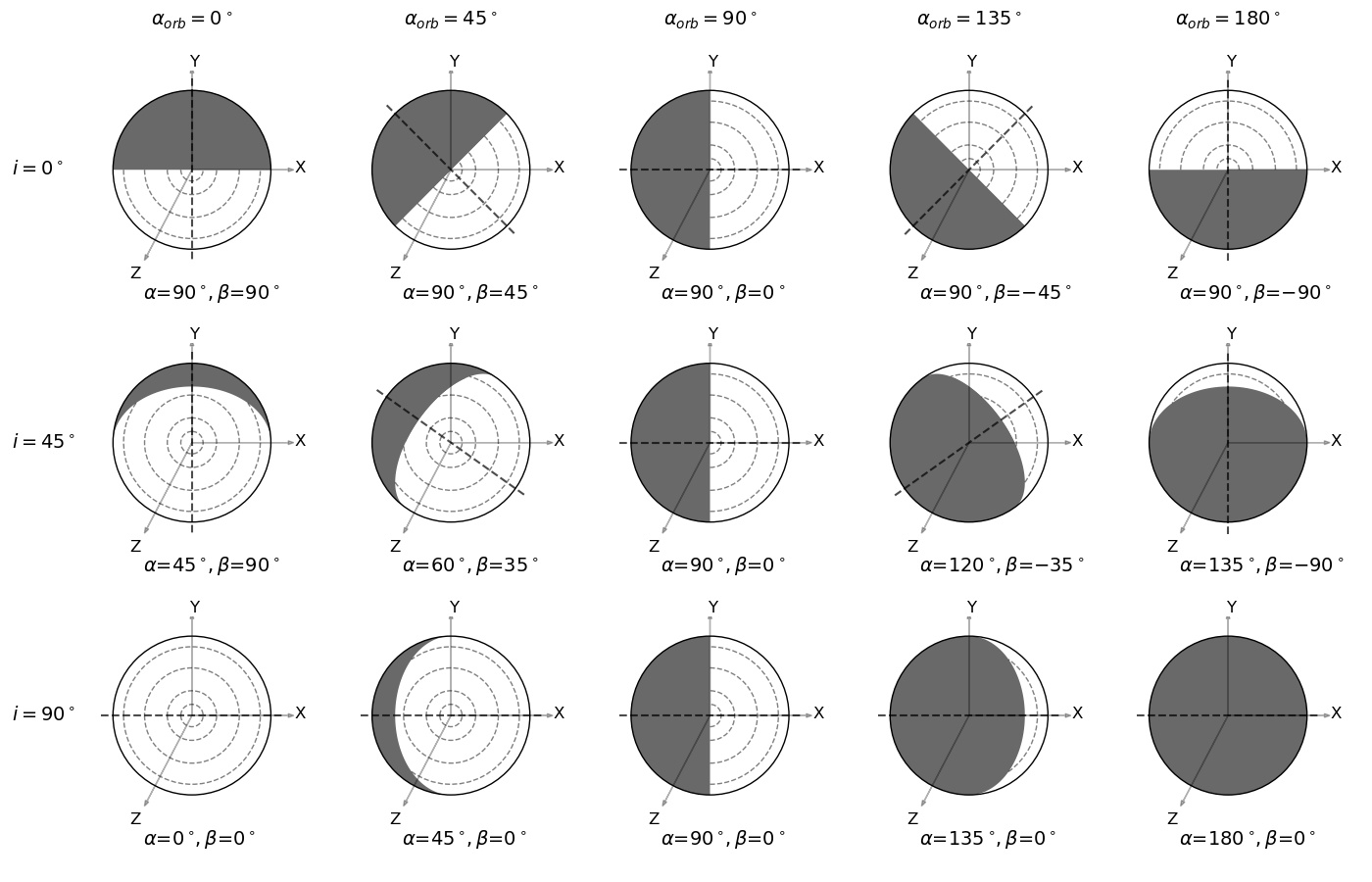}
\caption{The dayside (illuminated) and the night-side of a planet for different inclination angle ($i$) and different orbital phase ($\alpha_{orb}$). The corresponding phase angles ($\alpha$) and the angles between the planetary scattering plane and the XZ-plane ($\beta$) are shown. The planetary scattering planes for different $i$ and $\alpha_{orb}$ are shown with dashed lines.
\label{fig:daynight-all}}
\end{figure}

For an orbital inclination angle $i$ and orbital phase $\alpha_{orb}$, the angle between the planetary scattering plane and the optical plane can be written as (see Figure~\ref{fig:daynight-all}),
\begin{equation} \label{eq:beta}
\beta = \arccos{\frac{\sin \alpha_{orb}}{\sqrt{\sin^2\alpha_{orb}+\cos^2\alpha_{orb}\cos^2 i}}}
\end{equation}

Hence, the reflected intensity vector with respect to the optical plane of observation at \Mphi ~on the disk can be expressed as,
\begin{equation} \label{eq:iobs}
\vec{I}(M,\Phi) = \vec{L}(\beta)\vec{I_{scat}}(M,\Phi)
\end{equation}

To get the disk-integrated flux vector from the intensity vectors we follow a numerical technique developed by us described in the next section.

\section{Integration over the Planetary disk} \label{sec:diskint}

The disk-integrated flux vector \mvec{F} defined with respect to the optical plane of observation can be obtained by,
\begin{equation} \label{eq:iqudisk1}
\vec{F} = \int_{\rightmoon} \vec{I(M,\Phi})~ d\omega
\end{equation}

We first integrate \mvec{I}($M,\Phi$) along the concentric circular segments within the dayside at each $M$ defined on the grids of disk locations and then integrate over $M$. For each grid-value of $M$, the integration boundaries on $\Phi$ are given by $\pm\Phi_{term}(M,\alpha)$, which coincide with the terminator, i.e. the border between the dayside and the night-side on the disk facing the observer. $\Phi_{term}(M,\alpha)$, which is a function of both $M$ and phase angle $\alpha$, can be expressed as,
\begin{equation} \label{eq:phiterm}
\Phi_{term}(M,\alpha)= 
\begin{cases}
   \arccos{-\frac{M~\cot \alpha}{\sqrt{1-M^2}}},& \text{if } M \leq |\sin \alpha|\\
    \pi,              & \text{if } M > |\sin \alpha|~ \text{and } \alpha < \pi/2\\
    0, & \text{otherwise}
\end{cases}
\end{equation}

Thus Equation~\ref{eq:iqudisk1} can be rewritten as,
\begin{equation} \label{eq:iqudisk2}
\vec{F} = \frac{R_P^2}{D^2}\int_0^1\int_{-\Phi_{term}}^{\Phi_{term}} \vec{I(M,\Phi})~d\Phi M dM
\end{equation}

We perform this integration numerically. We choose 64 Gaussian points for the values of $M$ and $\mu$ and integrate over $M$ or $\mu$ using Gauss's quadrature formula. On the other hand, we choose linearly spaced values for $\Phi$ and $\phi$ and integrate over them using Simpson's rule.

The flux vector \mvec{F} can be expressed in terms of its component parameters as,
\begin{equation} \label{eq:fvec}
\vec{F} = \vector{F}{Q_{disk}}{U_{disk}}{V_{disk}},
\end{equation}

where $F$ is the reflected scalar flux, $Q_{disk}$ and $U_{disk}$ denote the disk-inetgrated linear polarizations and $V_{disk}$ denotes the disk-inetgarted circular polarization. As we assume the incident starlight to be unpolarized and we focus only on the linear polarization, we can set $V(M,\Phi)=0$. If we block the unpolarized light from the host star, the total disk-integrated degree of linear polarization can be expressed as,
\begin{equation} \label{eq:pdisk}
P_{disk} = \frac{\sqrt{Q_{disk}^2+U_{disk}^2}}{F}
\end{equation}

Taking $\frac{R_P^2}{D^2}=1$ in Equation~\ref{eq:iqudisk2}, we get the reflected flux at the TOA of the planet and we denote this flux by $F$. On the other hand, the inclusion of this factor provides the reflected flux to the observer at a distance $D$. We define the flux at the observer by $F_{obs}$. We calculate $F$ (or, $F_{obs}$), $Q_{disk}$, $U_{disk}$, and $P_{disk}$ for different atmospheric conditions and for different values of inclination angle and orbital phase as elaborated in the following sections.

\section{Benchmark Analysis} \label{sec:benchmark}

\begin{figure}[!ht]
\centering
\includegraphics[scale=0.33,angle=0]{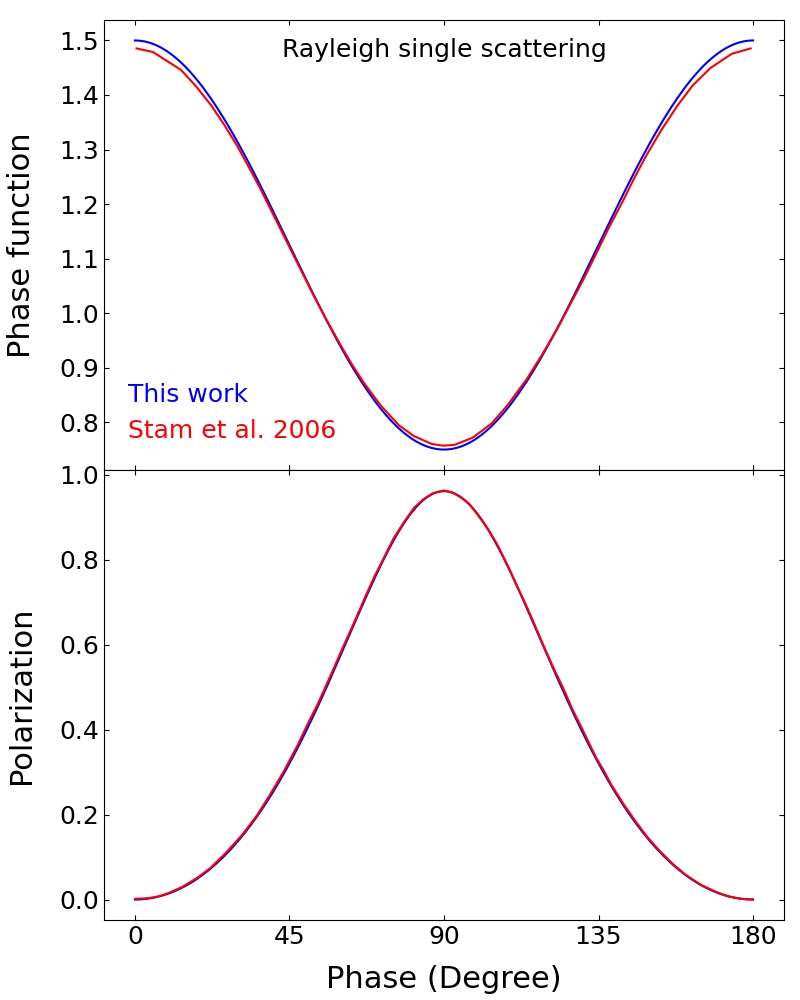}
\caption{Models for the phase function ($P_{11}$) and polarization for a single scattering ($P_{12}$/$P_{11}$) due to gas molecules (Rayleigh scattering) (see Section~\ref{sec:benchmark}) calculated by using our numerical code and that presented by \cite{stam06}.
\label{fig:benchmark-raysingle}}
\end{figure}

\begin{figure}
\centering
\includegraphics[scale=0.33,angle=0]{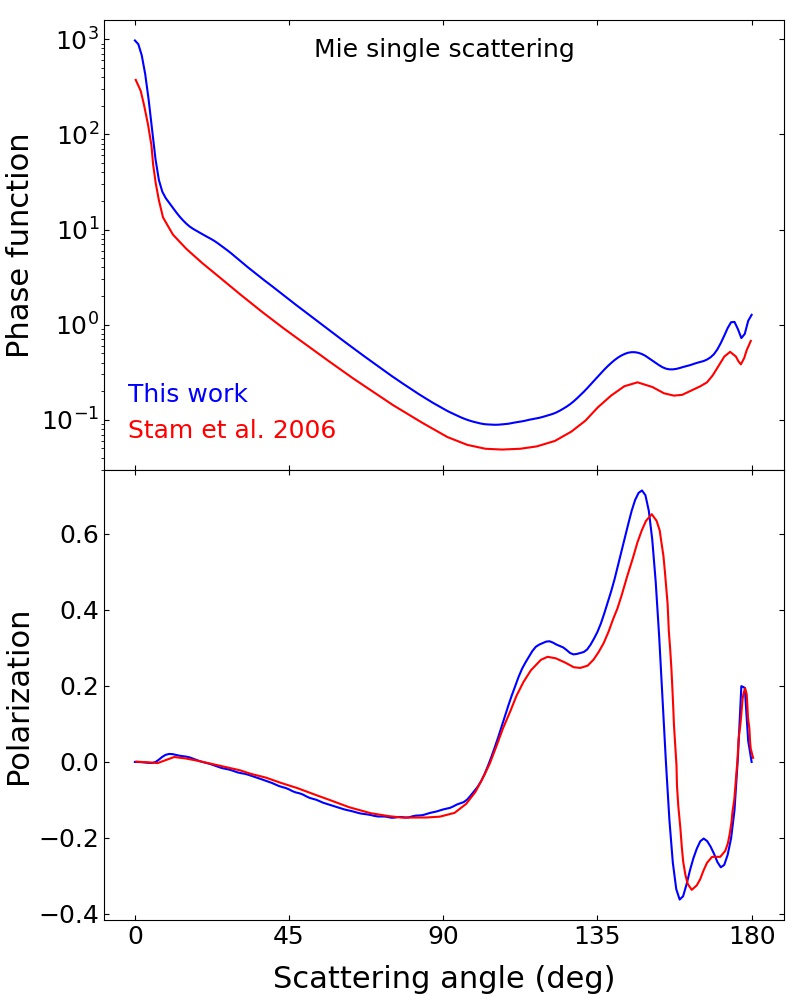}
\caption{Same as Figure~\ref{fig:benchmark-raysingle}, except, the scattering is due to cloud particles (Mie scattering).
\label{fig:benchmark-miesingle}}
\end{figure}

As a part of our benchmark analysis, we compare some of the results derived by using our numerical package with the results presented in some of the published work. First, we compare the scalar phase function and the polarization for a single scattering with those calculated by \cite{stam06} for both Rayleigh scattering and Mie scattering at 0.55 $\mu$m wavelength. In the case of Rayleigh scattering, we adopted a depolarization factor of 0.02 as assumed by \cite{stam06}. In the case of Mie scattering, we considered the cloud particles of refractive index equal to 1.33 (imaginary part is 0) with a size distribution having the mean at 2.2 $\mu$m as assumed by \cite{stam06,rooij84}. However, we assume a log-normal size distribution of the particles \citep[e.g.,][]{saumon00, sengupta20, chakrabarty20} which is different from what \cite{stam06, rooij84} assumed. Figure~\ref{fig:benchmark-raysingle} and Figure~\ref{fig:benchmark-miesingle} demonstrate the comparisons with the results presented by \cite{stam06} regarding the variations in phase function and polarization with the varying scattering angle. Clearly, the figures show that our calculations are in close agreement with that obtained by \cite{stam06}.

\begin{figure}
\centering
\includegraphics[scale=0.33,angle=0]{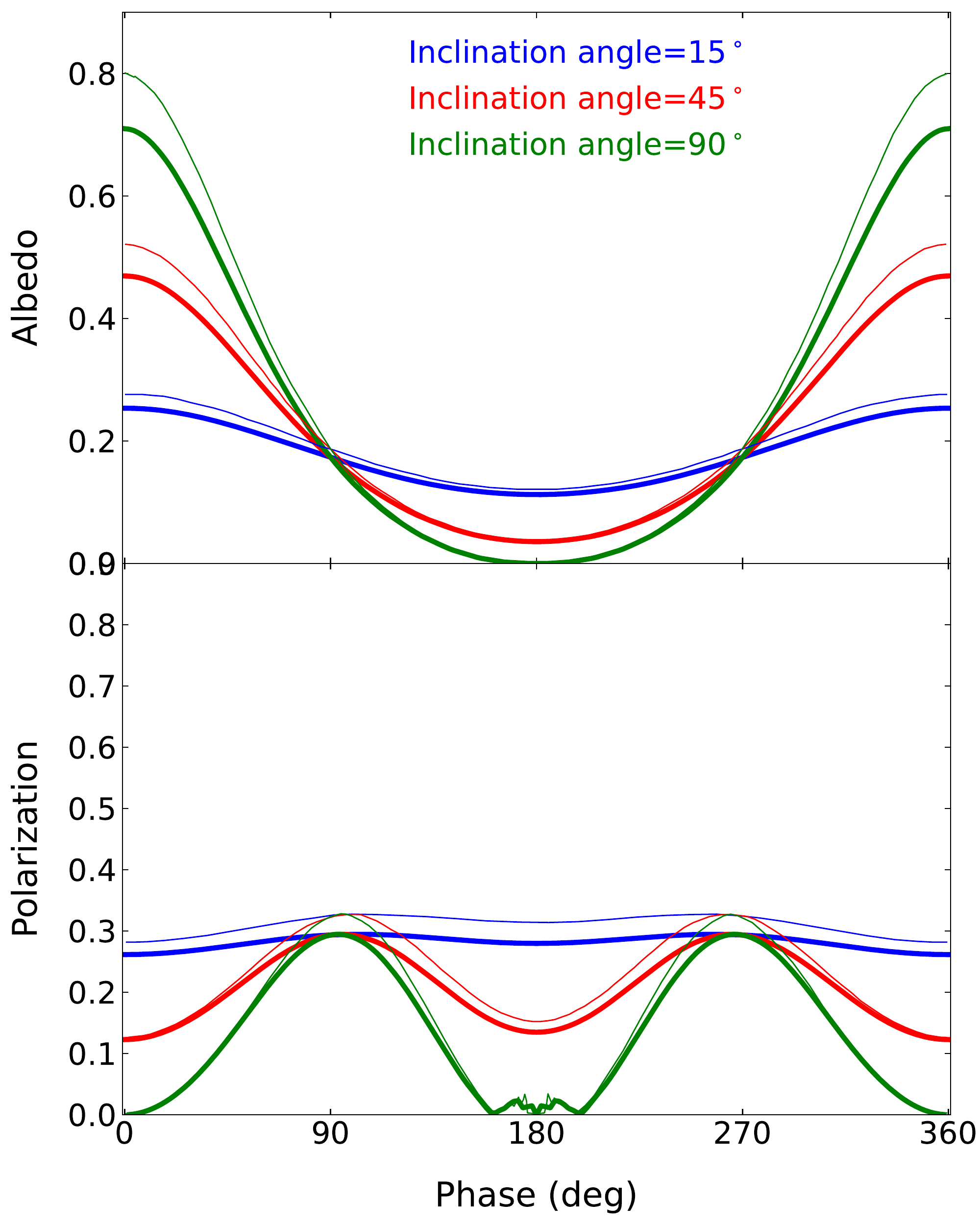}
\caption{Models for the albedo ($F(\alpha)/F_0$) and the polarization ($P_{disk}$) for a semi-infinite atmosphere obeying the principle of invariance with conservative Rayleigh scattering ($\omega=1$) calculated in the present work (thick lines) and that presented by \cite{madhusudhan12} (thin lines).
\label{fig:benchmark-multi}}
\end{figure}

Next, we benchmark our calculations of the disk-integrated total degree of polarization with the calculations of \cite{madhusudhan12}, considering conservative Rayleigh multiple scattering. \cite{madhusudhan12} assume a semi-infinite atmosphere obeying the principle of invariance \citep[e.g.,][]{chandrasekhar60}. To achieve this, we keep on adding optically thin ($d\tau \lesssim 1$) layers until the albedo and the polarization stop changing with the further addition of layers. At this point,  the atmosphere obeys the principle of invariance. For the conservative case, we set the single scattering albedo ($\omega$) equal to 1. Figure~\ref{fig:benchmark-multi} shows the comparison. Both the models are in close accord.
%\clearpage

We have also performed a benchmark analysis of the emergent flux from the planet and we discuss it in the next section.

\section{ The Planetary Atmosphere Model} \label{sec:atmosphere}

\subsection{Model for a Cloud-free Atmosphere} \label{sec:atmosphere/cloudless}

Our numerical code allows us to model the atmospheres of reflecting exoplanets with realistic predictions of the observable spectra and the phase curves. This is achieved in a few steps. The first step involves calculations of the pressure-temperature (P-T) profile of the atmosphere by using the Fortran\footnote{\url{http://cdsarzontally c.u-strasbg.fr/viz-bin/qcat?J/A+A/574/A35}} implementation of the analytical models of non-Grey irradiated planets presented by \cite{parmentier15} which is based on the Rosseland opacities presented in \cite{freedman08} and the functional form of Rosseland opacity derived by \cite{valencia13}. The details of the implementation of this code in our model can be found in \cite{sengupta20, chakrabarty20}. The internal temperature of the planet is assumed to be 200K for all the cases.

The atmospheric composition can be fed to our code either directly or from any database. Here, we adopt solar metallicity and solar system abundance for the atoms and the molecules present in the atmospheres. The atomic and molecular species, that we choose as absorbers and scatterers, are H${\rm _2}$, He, VO, TiO, CO${\rm _2}$, H$\rm_2$O, CH$_4$, CO, Na, and K. As we consider the atmospheres of the hot Jupiters, H${\rm _2}$ acts as the dominant scatterer when considering Rayleigh scattering. The mass-fractions for all these atoms and molecules have been calculated using the abundance database given in the open-source package \texttt{Exo-Transmit}\footnote{\url{https://github.com/elizakempton/Exo_Transmit}} provided by \cite{kempton17}. In the present work, we have considered the rain-out condensation model. Our numerical code also allows inputs from any opacity database in order to calculate the absorption and scattering coefficients. Here, we have used the opacity database \citep{lupu14,freedman14,freedman08} from the same package, \texttt{Exo-Transmit} as explained in \cite{sengupta20, chakrabarty20}. Both the abundance and the opacity databases are available over a broad grid of pressure and temperature from which the required opacities are calculated by interpolation for a specific P-T profile. 

\subsection{Benchmark Analysis of the Emergent Planetary Flux} 

\begin{figure}
\centering
\includegraphics[scale=0.33,angle=0]{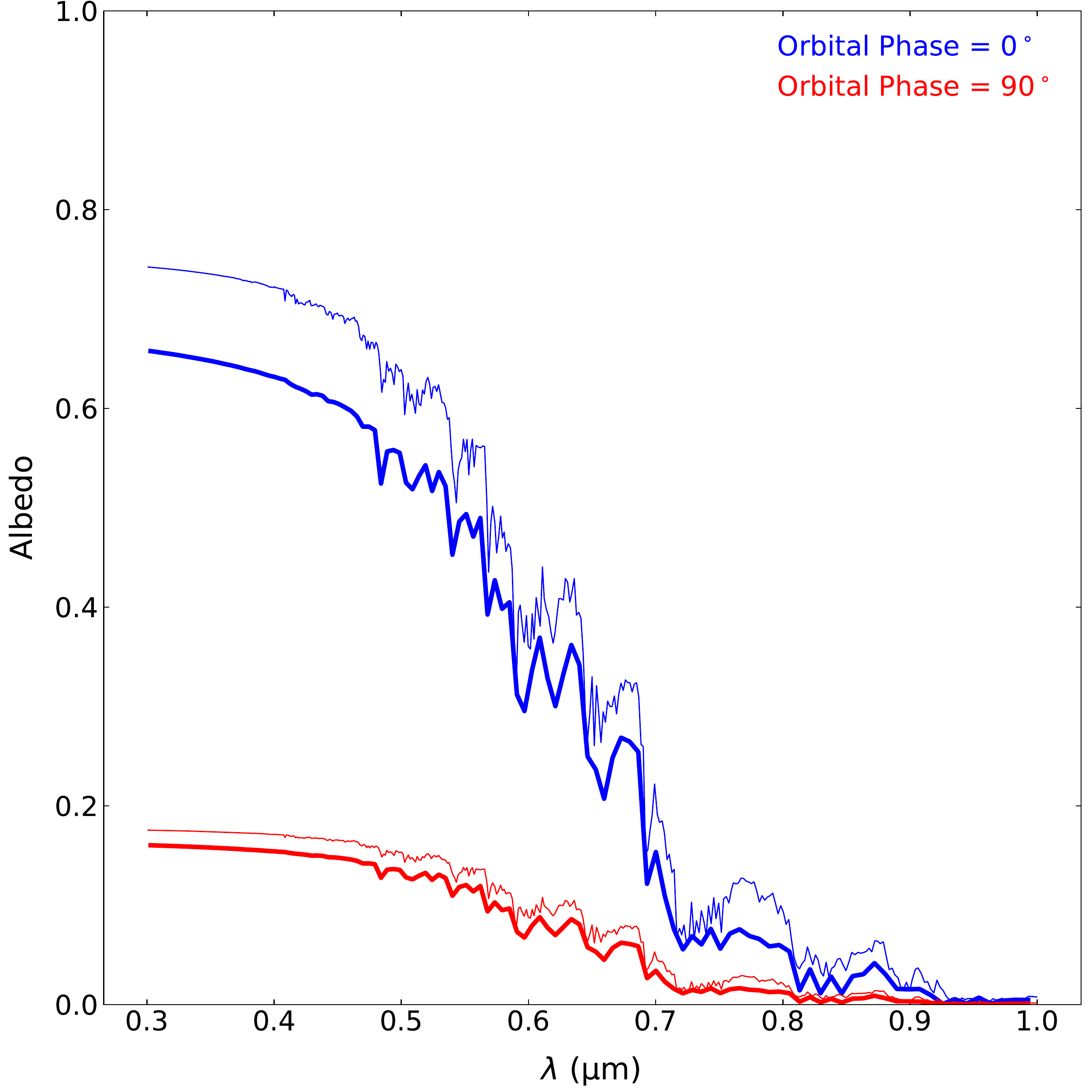}
\caption{The albedo ($F(\alpha)/F_0$) of a hot Jupiter with $T_{eq}=1200$ K and $g=30$ ms$^{-2}$ (see Section~\ref{sec:benchmark}) calculated by using our numerical code (thick lines) and by using the open-source code \texttt{Picaso} developed by \cite{batalha19} (thin lines) at two different orbital phases, orbital inclination angle being fixed at 90$^\circ$.
\label{fig:benchmark-flux}}
\end{figure}

We compare the reflection spectra ($F(\lambda)$) calculated for phase angle equal to 0\deg\!\!, and 90\deg and for an inclination angle of 90$^\circ$ over a wavelength ranging between 0.3 ${\rm\mu}$m and 1 ${\rm \mu}$m  with those calculated by using the  numerical code \texttt{picaso}\footnote{\url{https://github.com/natashabatalha/picaso}} developed by \cite{batalha19} and available in public domain. For this purpose, we consider a cloud-free atmosphere with H${\rm _2}$, He, VO, TiO, CO${\rm _2}$, H$\rm_2$O, CH$_4$, CO, NH$\rm_3$, N$\rm_2$, and PH$\rm_3$. Figure~\ref{fig:benchmark-flux} shows that the two spectra are in close agreement, especially at wavelengths greater than 0.5 $\mu$m and they exhibit similar spectral features over wavelength. This demonstrates the accuracy of our calculations at all the wavelength points in the optical region. The difference between the two albedo models, especially in the lower wavelength region, can be attributed to some difference in the numerical techniques adopted as well as in the estimation of the atomic and molecular opacity.

\subsection{Effects of the Clouds} \label{sec:atmosphere/cloudy}

Water cannot condense at high temperature prevailing in the atmospheres of hot Jupiters and instead, iron or silicate condensates are likely to form the clouds in the upper atmospheres of the hot Jupiters \citep[e.g.,][]{chakrabarty20, sengupta20, lodders10, ackerman01}. Here, we calculate the effect of clouds on the albedo and polarization of the hot Jupiters by considering Forsterite (Mg$\rm_2$SiO$\rm_4$) as the dominant constituent of the clouds. We assume that the cloud particles are located vertically within a single region of the atmosphere bound by a base and a deck. Within this region, the particles are assumed to be distributed evenly horizontally, i.e. the cloud density is homogeneous along a stratified layer of the atmosphere. The particles follow a vertical distribution of density given by \citep{saumon00, sengupta20, chakrabarty20} the relation $n(P)=n_0P/P_0$, where, n(P) is the number density of the cloud particles at a pressure level P, and $n_0$ is the number density at the reference pressure level $P_0$. We set $P_0$ at 1 bar. We consider a lognormal distribution of the size (diameter) of the particles with the mean diameter ($d_0$) as a free parameter \citep{ackerman01, sengupta20, chakrabarty20}. As $d_0$ tends to zero, light scattered by these particles tends to follow the Rayleigh theory of scattering. This is known to be representative of the haze in the atmosphere of Titan (see, e.g., \cite{saumon00}).
 
We have calculated the extinction coefficients, the scattering coefficients, and the phase matrices of the condensate cloud following the Mie theory of scattering \citep[][etc.]{vandehulst57,hansen74,fowler83,bohren83,sengupta20}. The approach we follow in developing our code allows fast calculation of the extinction and scattering cross-sections due to Mie scattering as well as the Mie phase matrices as functions of \muphi.

\begin{figure}
\centering
\includegraphics[scale=0.3,angle=0]{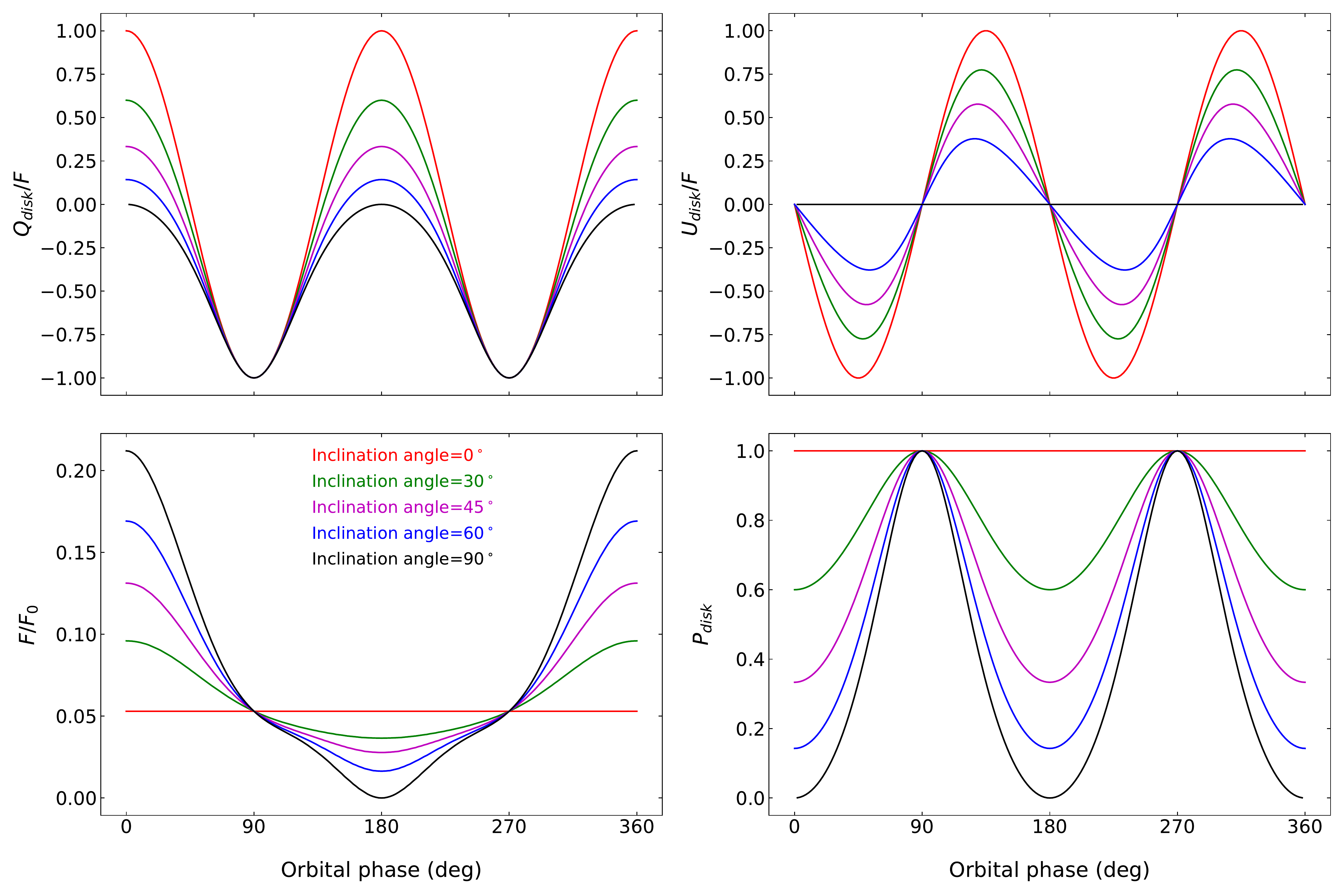}
\caption{The albedo ($F(\alpha)/F_0$) and the disk-integrated polarization ($P_{disk}$) phase curves for the cloud-free atmosphere of type-i (see Section~\ref{sec:effects-multi}) for different inclination angle at $\lambda=0.55$ ${\rm \mu}$m, considering only single scattering of the incident light in the atmosphere.
\label{fig:atmi-single}}
\end{figure}

\begin{figure}
\centering
\includegraphics[scale=0.3,angle=0]{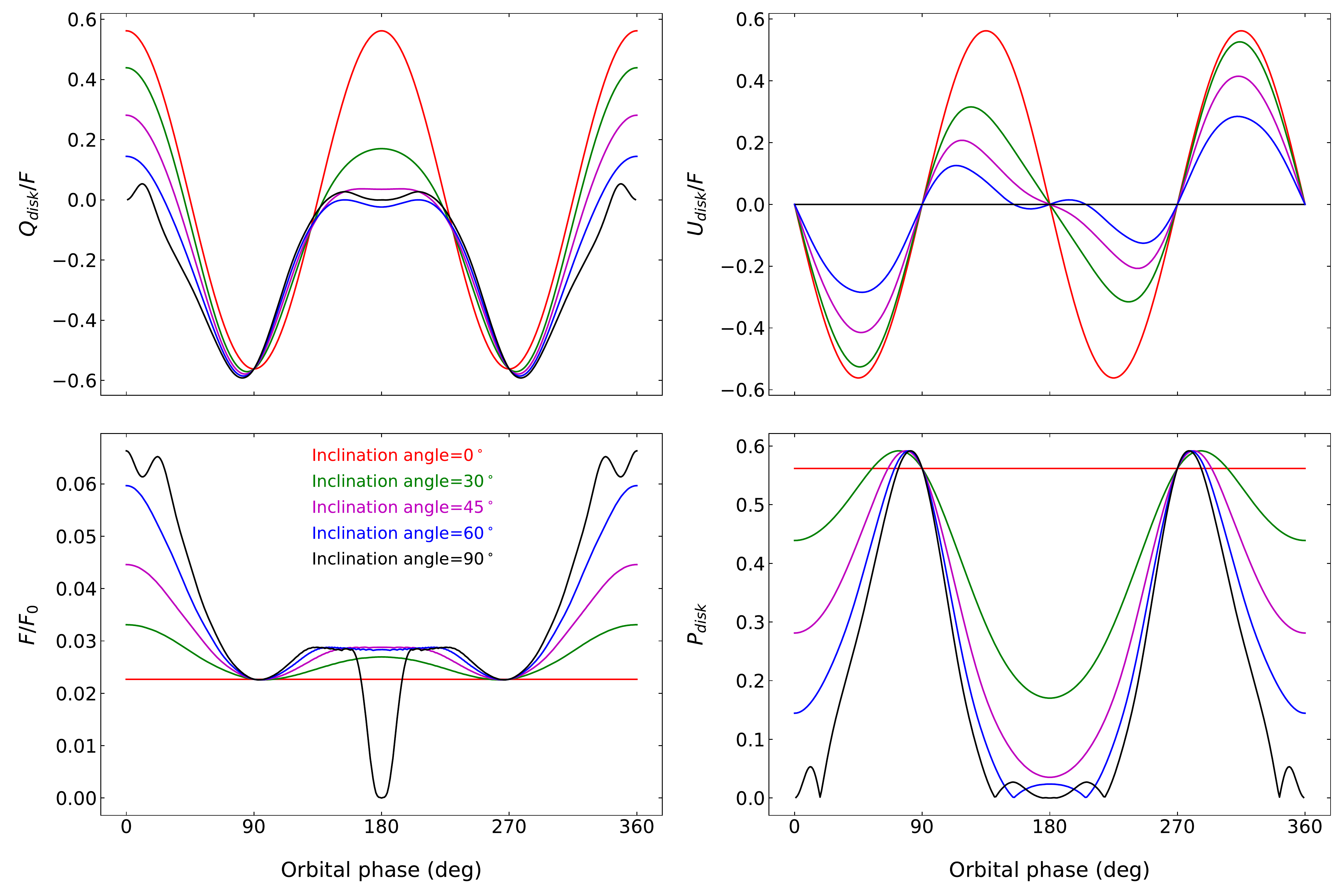}
\caption{The albedo ($F(\alpha)/F_0$) and the disk-integrated polarization ($P_{disk}$) phase curves for the cloudy atmosphere of type-ii (see Section~\ref{sec:effects-multi}) for different inclination angle at $\lambda=0.55$ ${\rm \mu}$m, considering only single scattering of the incident light in the atmosphere.
\label{fig:atmii-single}}
\end{figure}

\begin{figure}
\centering
\includegraphics[scale=0.5,angle=0]{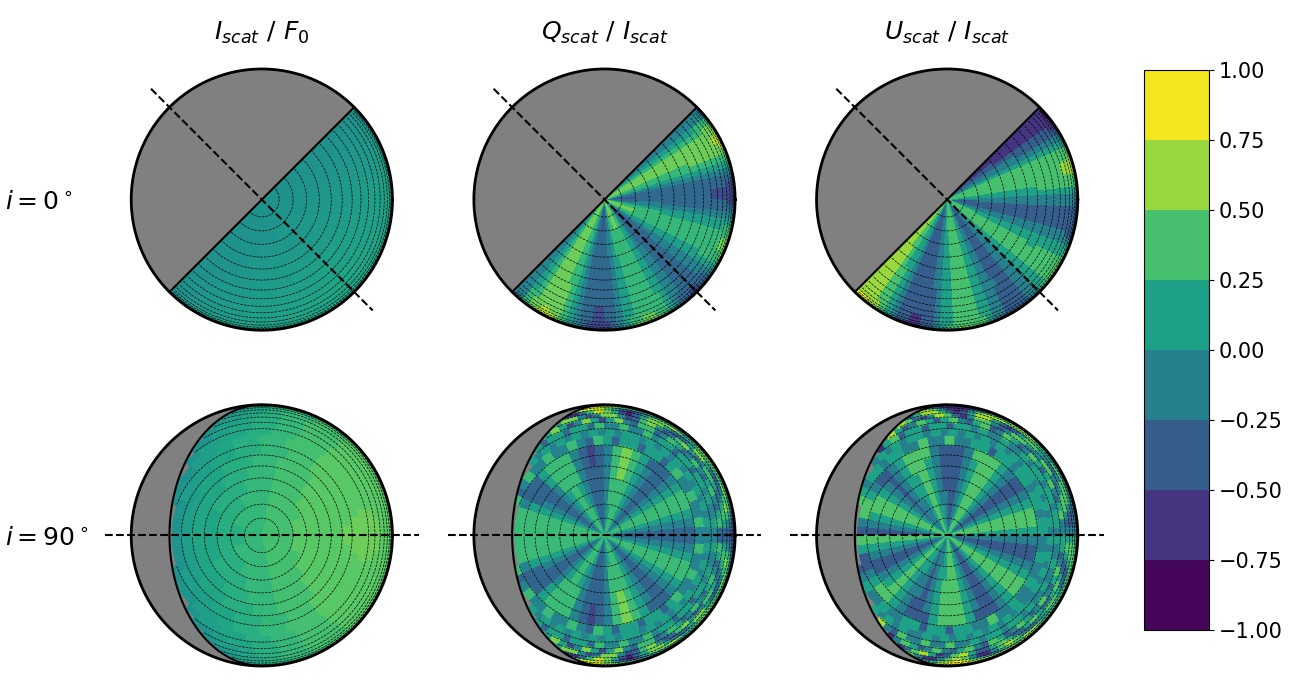}
\caption{The normalized disc-resolved intensity vector components for the cloud-free atmosphere of type-i (see Section~\ref{sec:effects-multi}) for two different inclination angles and $\alpha_{orb}=45^\circ$ at $\lambda=0.55$ ${\rm \mu}$m, considering both single scattering of the incident light and multiple scattering of the internal radiations in the atmosphere.
\label{fig:atmi-multi-iscat}}
\end{figure}

\begin{figure}
\centering
\includegraphics[scale=0.5,angle=0]{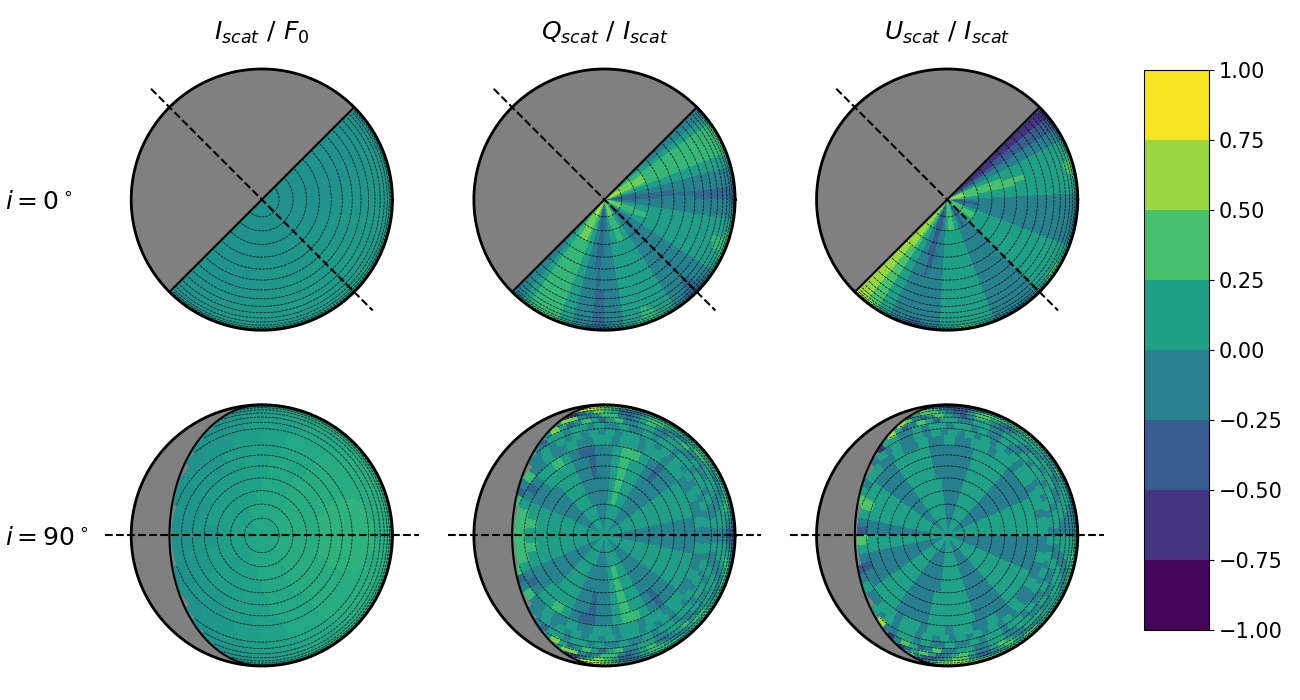}
\caption{The normalized disc-resolved intensity vector components for the cloudy atmosphere of type-ii (see Section~\ref{sec:effects-multi}) for two different inclination angles and $\alpha_{orb}=45^\circ$ at $\lambda=0.55$ ${\rm \mu}$m, considering both single scattering of the incident light and multiple scattering of the internal radiations in the atmosphere.
\label{fig:atmii-multi-iscat}}
\end{figure}

\begin{figure}
\centering
\includegraphics[scale=0.3,angle=0]{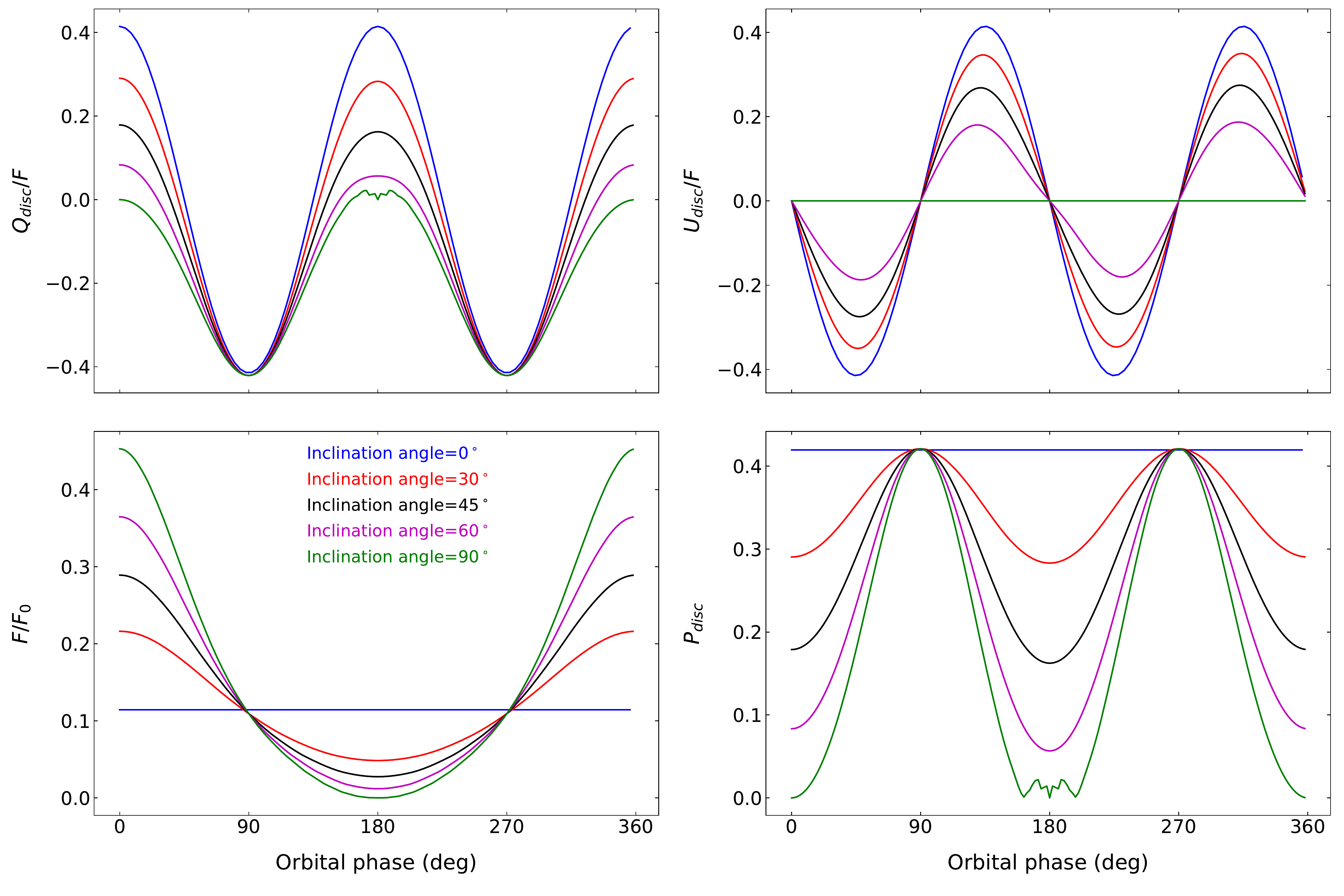}
\caption{The albedo ($F(\alpha)/F_0$) and the disk-integrated polarization ($P_{disk}$) phase curves for the cloud-free atmosphere of type-i (see Section~\ref{sec:effects-multi}) for different inclination angle at $\lambda=0.55$ ${\rm \mu}$m, considering both single scattering of the incident light and multiple scattering of the internal radiations in the atmosphere.
\label{fig:atmi-multi}}
\end{figure}

\begin{figure}
\centering
\includegraphics[scale=0.3,angle=0]{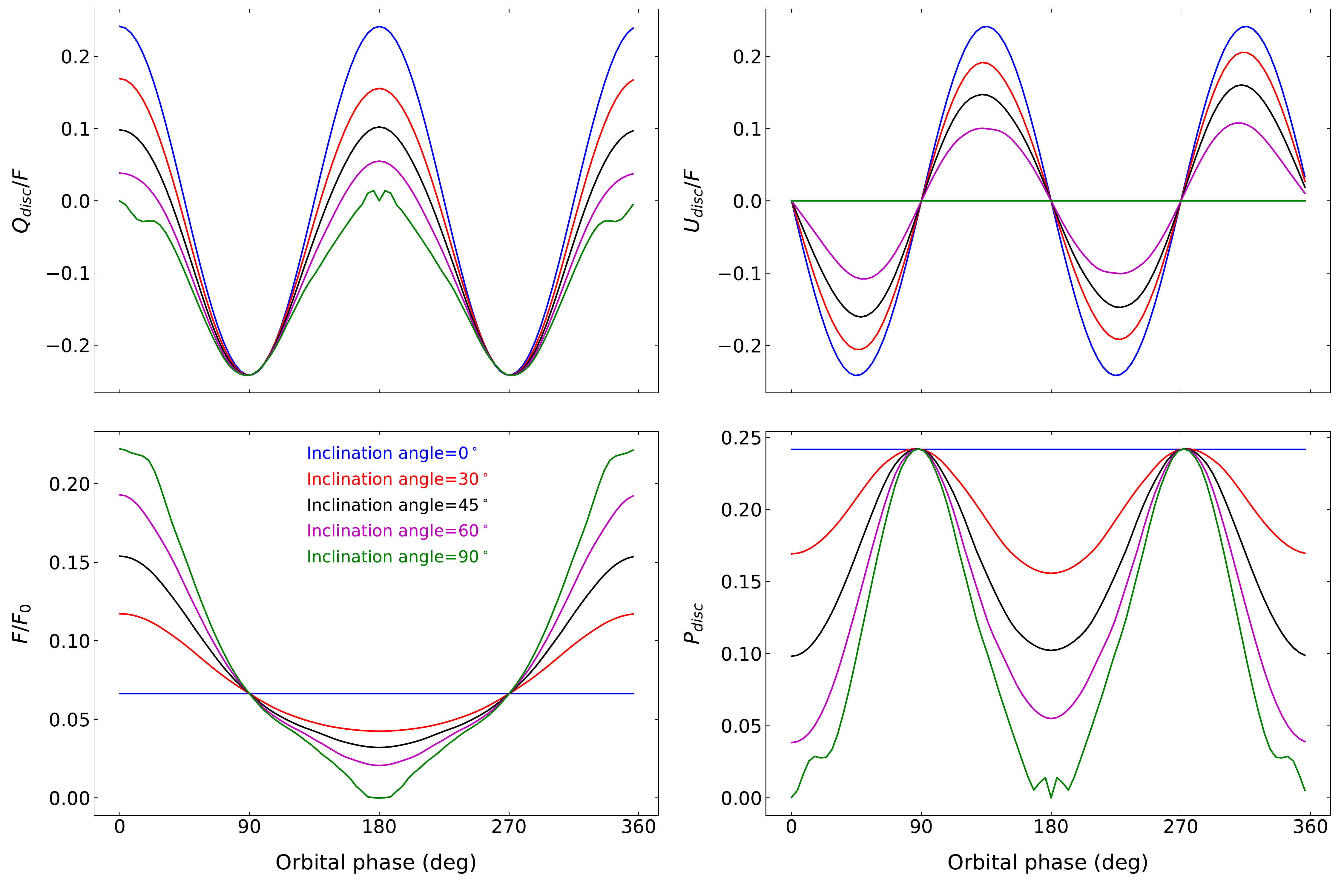}
\caption{The albedo ($F(\alpha)/F_0$) and the disk-integrated polarization ($P_{disk}$) phase curves for the cloudy atmosphere of type-ii (see Section~\ref{sec:effects-multi}) for different inclination angle at $\lambda=0.55$ ${\rm \mu}$m, considering both single scattering of the incident light and multiple scattering of the internal radiations in the atmosphere.
\label{fig:atmii-multi}}
\end{figure}

\section{Single Scattering Vs. Multiple Scattering} \label{sec:effects-multi}

When we consider only the effect of the single scattering of the incident light in the atmosphere of a planet, the degree of polarization of the reflected light is somewhat overestimated, whereas, the reflected flux (scalar) is underestimated. Multiple scattering of the internal radiations increases the amount of light scattered in the direction of the observer and at the same time, causes an effect of depolarization of the reflected light \citep[e.g.,][]{stam06, sengupta08, madhusudhan12, bailey18, batalha19}. In our calculations, we consider two types of atmospheres with $g=30$ ms$^{-2}$, viz. i) cloud-free atmosphere with chemical species such as H${\rm _2}$, He, VO, TiO, CO${\rm _2}$, H$\rm_2$O, CH$_4$, and CO; ii) the same atmosphere but with a high-altitude thin layer of clouds with $n_0 = 500$ cm$^{-3}$ and mean grain diameter equal to 1 $\mu$m, and the deck and base are defined at pressure levels of 200 Pa and 2000 Pa respectively.

We present the normalized components of the disk-resolved intensity vector \mvec{I_{scat}}, viz. $I_{scat}/F_0$, $Q_{scat}/I_{scat}$, and $U_{scat}/I_{Scat}$ in Figure~\ref{fig:atmi-multi-iscat}-\ref{fig:atmii-multi-iscat} for the atmopsheres of both type-i and type-ii, considering multiple scattering of light, for an orbital phase ($\alpha_{orb}$) of 45\deg and inclination angles ($i$) equal to 0\deg and 90\deg\!\!\!. However, we cannot observe these reflected intensity and polarization patterns over the disk of an exoplanet yet as we cannot resolve the disk of a close-in exoplanet with the current technology. We can only observe the disk-integrated flux and polarization from the exoplanets. The disk-integrated polarizations $Q_{disk}$, $U_{disk}$, and $P_{disk}$ and the albedo $F/F_0$ calculated for both the cases of single scattering and multiple scattering are shown in Figure~\ref{fig:atmi-single}-\ref{fig:atmii-multi}, at a wavelength of 0.55 $\mu$m.

\begin{figure}[!ht]
\centering
\includegraphics[scale=0.33,angle=0]{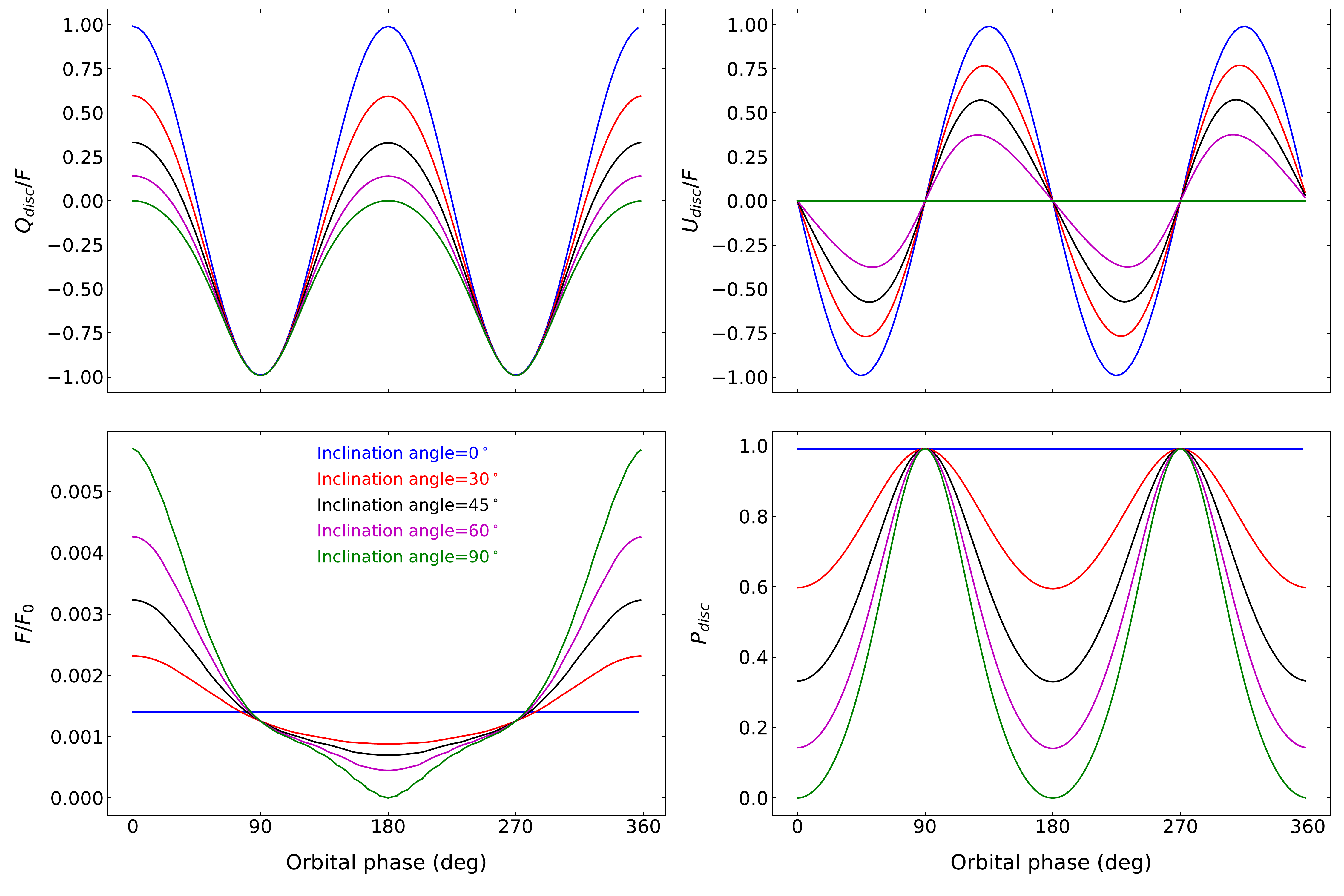}
\caption{Effect of Na and K on the albedo and the disk-integrated polarization phase curves at $\lambda=0.55$ ${\rm \mu}$m for an orbital inclination of 90\deg\!\!, considering both single scattering of the incident light and multiple scattering of the internal radiations in the cloud-free atmosphere of type-i (see Section~\ref{sec:effects-multi}).
\label{fig:NaK-atmi}}
\end{figure}

We also present the effect of the atomic absorbers, viz. Na and K, on the calculated phase curves of albedo and polarization. For this study, we consider the cloud-free (type-i) atmosphere mentioned above. Figure~\ref{fig:NaK-atmi} shows the corresponding results.

\begin{figure}[!ht]
\centering
\includegraphics[scale=0.4,angle=0]{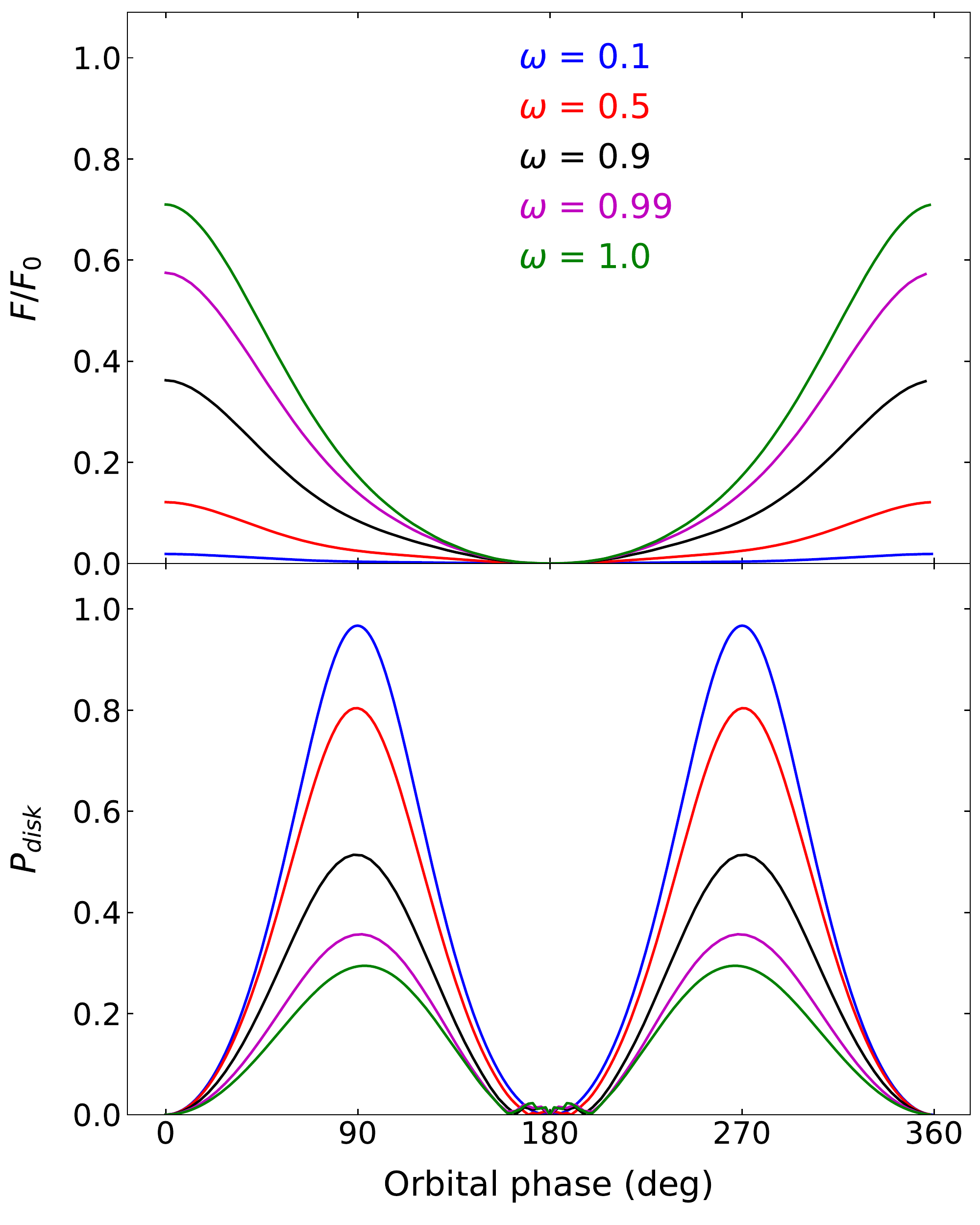}
\caption{Effect of single scattering albedo ($\omega$) on the albedo ($F(\alpha)/F_0$) and the disk-integrated polarization ($P_{disk}$) phase curves at $\lambda=0.55$ ${\rm \mu}$m for an orbital inclination of 90\deg\!\!, considering both single scattering of the incident light and multiple scattering of the internal radiations in the atmosphere.
\label{fig:omega}}
\end{figure}

One way to understand the effect of multiple scattering is to study the dependence of the albedo and the degree of polarization on the single scattering albedo ($\omega$). When considering only single scattering of the incident starlight, the albedo and the disk-integrated degree of polarization do not depend on $\omega$. However, as multiple scattering of the internal radiations is incorporated, one can find a strong dependence of the albedo and the polarization on $\omega$. Figure~\ref{fig:omega} shows the variation of $F/F_0$ and $P_{disk}$ with increasing $\omega$. This is in accord with the results (see Figure~4) of \cite{madhusudhan12}. For this study, we fix the total optical depth of the atmosphere by assuming a semi-infinite atmosphere obeying the principle of invariance as explained in \ref{sec:benchmark}.

\section{Models for HD 189733 \lowercase{b}}\label{sec:hd189733b}

\begin{figure}[!ht]
\centering
\includegraphics[scale=0.29,angle=0]{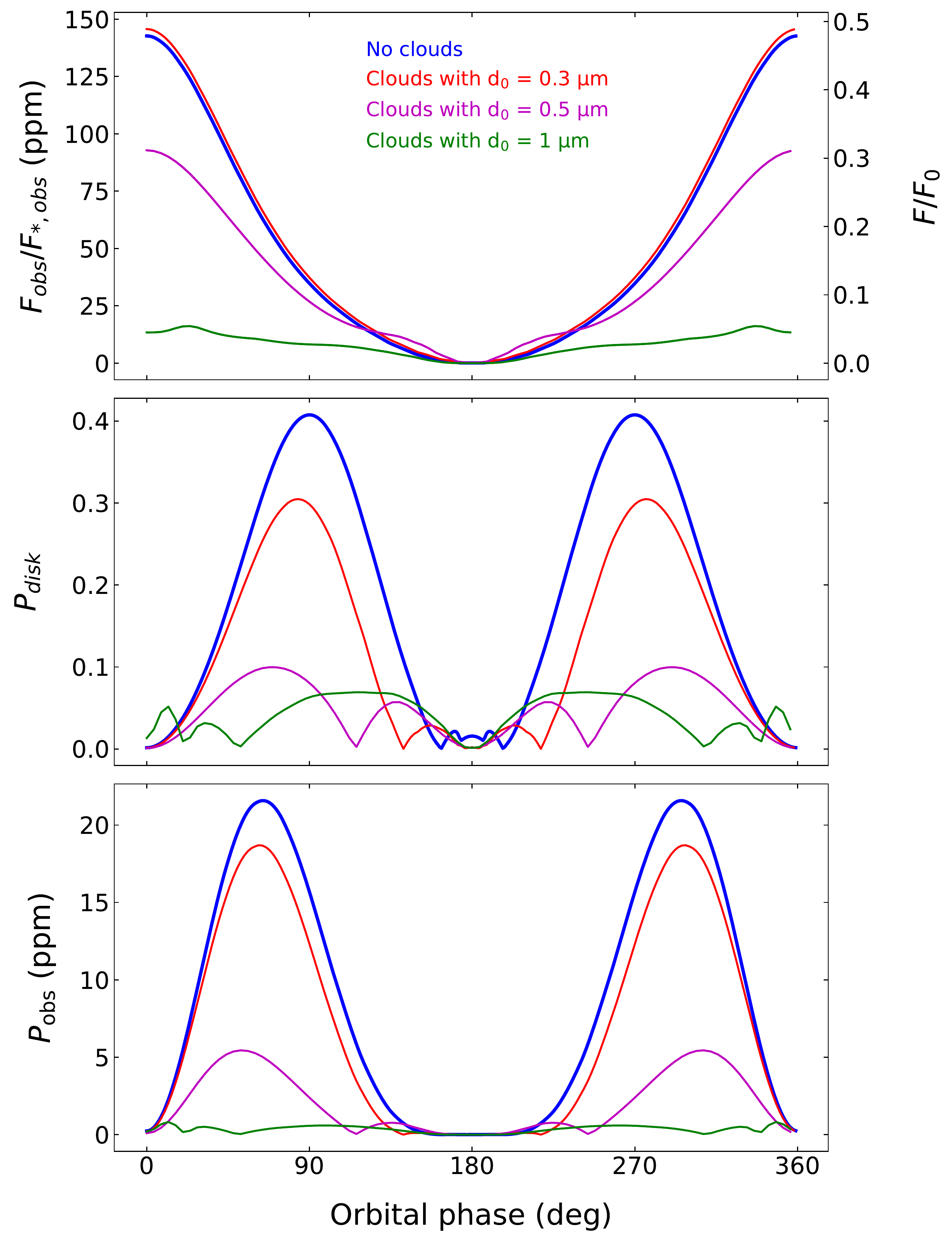}
\caption{Contrast of the reflected flux from HD 189733 b with respect to the host star flux ($F_{obs}/F_{*,obs} = F/F_*~\frac{R_P^2}{R_*^2}$) at the observer and the corresponding degree of polarization ($P_{obs}$) at 0.55 $\mu$m wavelength without including the atomic absorbers, viz. Na and K.
\label{fig:hd189733b}}
\end{figure}

\begin{figure}
\centering
\includegraphics[scale=0.29,angle=0]{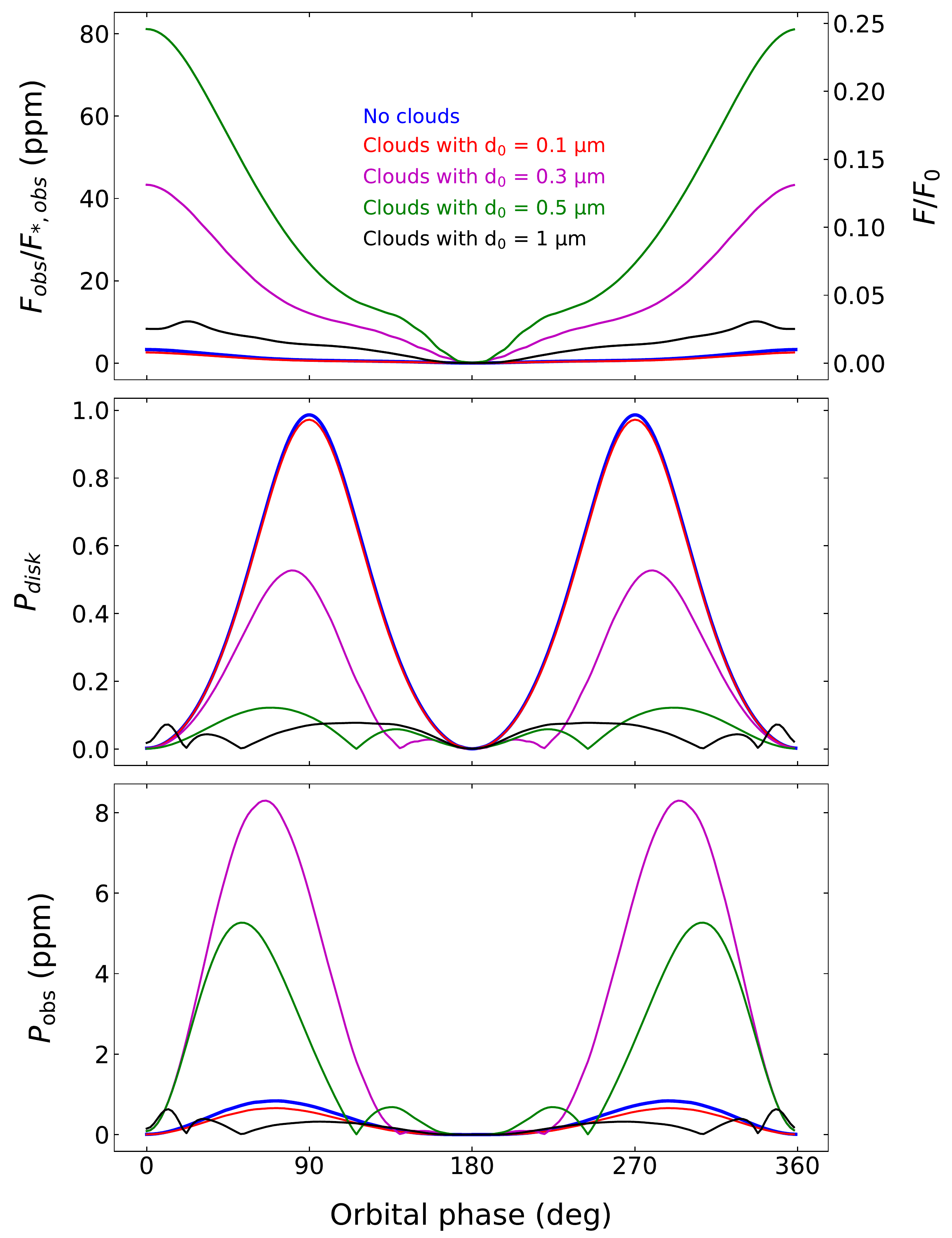}
\caption{Same as Figure~\ref{fig:hd189733b} but including the atomic absorbers, Na and K.
\label{fig:hd189733b-NaK}}
\end{figure}

\begin{figure}
\centering
\includegraphics[scale=0.29,angle=0]{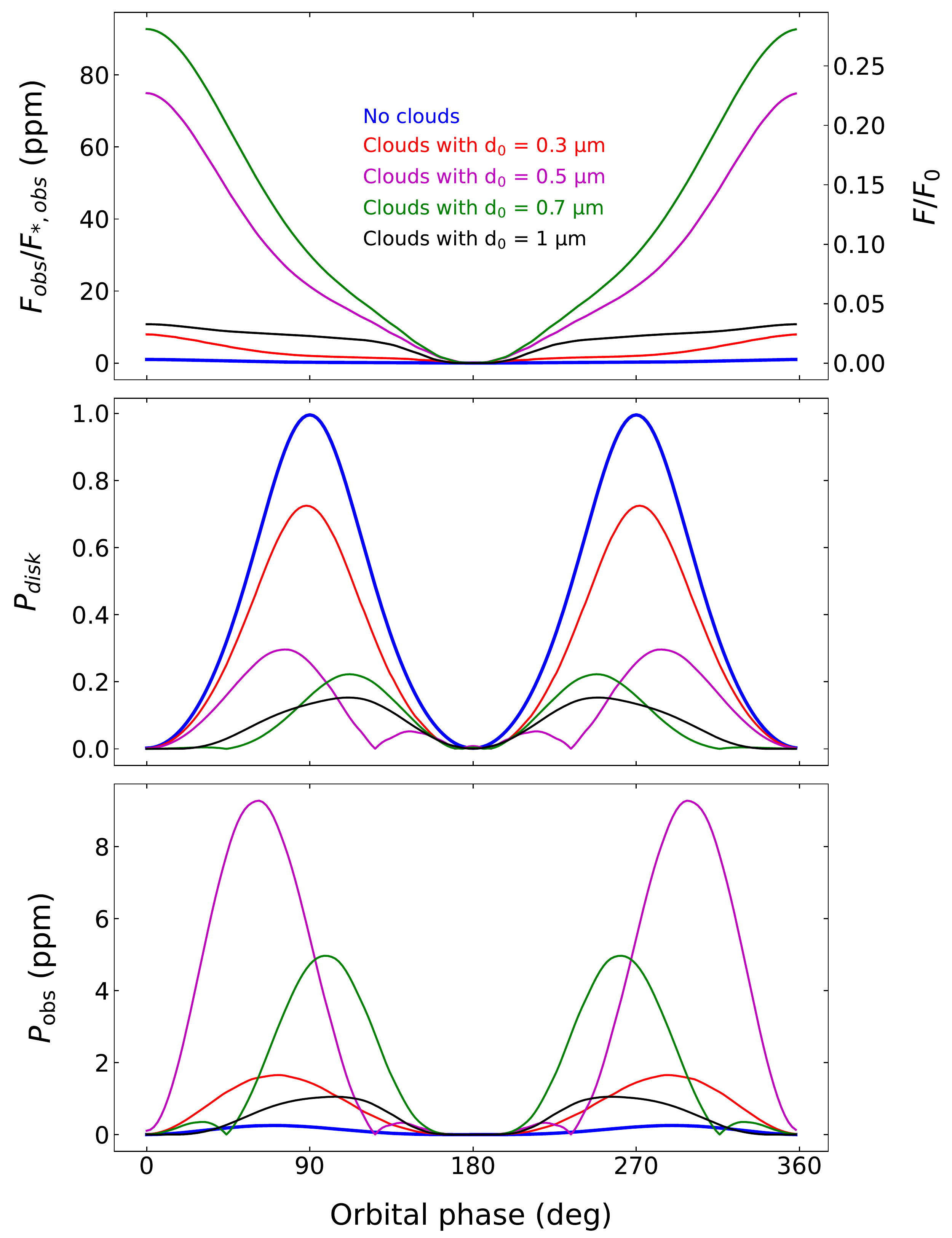}
\caption{Same as Figure~\ref{fig:hd189733b} but at 1.25 $\mu$m wavelength. At this wavelength the effect of Na and K is negligible.
\label{fig:hd189733bJ}}
\end{figure}

In order to guide the future polarimetric observation of hot Jupiters, we present models for the reflected flux and the disk-integrated polarization of the well-studied exoplanet HD 189733 b. We have adopted the properties of the planet and the host star from \cite{sengupta20} and \cite{addison19}. We consider both a cloud-free and a cloudy atmosphere for the planet. We adopted the same chemical species in the atmosphere as mentioned in Section~\ref{sec:atmosphere/cloudless}. Additionally, for the cloudy atmosphere, we consider different mean diameter ($d_0$) for the size-distribution of the cloud particles, ranging between 0.1 $\mu$m and 1 $\mu$m (see Figure~\ref{fig:hd189733b}-\ref{fig:hd189733bJ}). We have considered the cloud deck and base at the pressure levels of 200 Pa and 7500 Pa respectively and set $n_0 = 5000$ cm$^{-3}$. 

We have done the calculations at two different wavelengths, one in the visible wavelength region at 0.55 $\mu$m and the other in the near infrared region at 1.25 $\mu$m. Figure~\ref{fig:hd189733b} shows our models for an atmosphere without the atomic absorbers, viz. Na and K, at 0.55 $\mu$m wavelength. Low-resolution transmission spectroscopic observations by \cite{mccullough14, sing09, swain08}; etc. suggest a hazy atmosphere for this planet which suppresses the atomic absorption features in the optical region. However, high-resolution transmission spectroscopic observation by \cite{redfield08} shows features of absorption due to atomic Na. So, we also present our models with atomic absorbers such as Na and K (see Figure~\ref{fig:hd189733b-NaK}). We add these atoms in certain mass-fractions calculated from the equation of state database, corresponding to the metallicity equal to 1, provided in the package \texttt{Exo-Transmit} \citep{kempton17, lodders03}. 

At 1.25 $\mu$m wavelength, the absorption due to Na and K is absent. Hence, we present the models at this wavelength without considering Na and K (see Figure~\ref{fig:hd189733bJ}).

We represent the host star flux at the stellar surface and at the observer by $F_*$ and $F_{*,obs}$ respectively. Clearly, $$F_{*,obs} = F_*\frac{R_*^2}{D^2},$$ where $R_*$ denotes the radius of the host star. Figure~\ref{fig:hd189733b}-\ref{fig:hd189733bJ} show the reflection flux contrast $F_{obs}/F_{*,obs}$ and the degree of polarization $P_{obs}$ at the observer for atmosphere. As the planet cannot be resolved from the host star with the current technology, the total observed degree of polarization at the observer is calculated by,
\begin{equation} \label{eq:pobs}
P_{obs} = \frac{\sqrt{Q_{disk}^2+U_{disk}^2}}{F+F_*\frac{R_*^2}{R_P^2}}
\end{equation}
From Equation~\ref{eq:pobs}, the relations among $F_{obs}/F_{*,obs}$, $P_{obs}$, the disk-integrated polarization $P_{disk}$ (assuming the host star is blocked) and the albedo $F/F_0$ of the planet can be found to be,
\begin{equation} \label{eq:pobs-pdisk-albedo-contrast}
\begin{aligned}
\frac{F_{obs}}{F_{*,obs}} &= \frac{F}{F_0}\frac{R_P^2}{a^2}\\
P_{obs} &\approx P_{disk} \frac{F}{F_0}\frac{R_P^2}{a^2},
\end{aligned}
\end{equation}
where $a$ is the distance between the planet and its host star. Equation~\ref{eq:pobs-pdisk-albedo-contrast} allows us to calculate $F_{obs}/F_{*,obs}$ and $P_{obs}$ from $P_{disk}$ and the albedo ($F/F_0$) of the planet. All the calculations are done at $\lambda=0.55$ ${\rm \mu}$m.

\section{Results and Discussion} \label{sec:rd}

Our entire approach towards the present problem can be segregated into two steps. In the first step, we calculate the local reflection matrices at different locations on the disk following the solution of the vector radiative transfer equations using the discrete space theory \citep{peraiah73}. The second step involves the integration of the local reflection matrices over the disk using simple numerical techniques as explained in Section~\ref{sec:diskint}. In both these steps, we need to adjust the reference plane for the intensity vectors or the flux vector a number of times to get the final reflected flux vector along the plane of observation. All the calculations are done using the numerical code developed by us. Figure~\ref{fig:benchmark-raysingle}-\ref{fig:benchmark-flux} show the results of our benchmark analysis and demonstrate the validity of our derivations.

\subsection{Single Scattering and Multiple Scattering}

A simple and computationally inexpensive approach to solving the VRT equations is to consider only single scattering of the incident starlight at each layer of the atmosphere, but this approach only estimates a lower limit for the albedo ($F/F_0$) and an upper limit for $P_{disk}$. This approach is extremely useful in verifying the early observations claiming the detection of polarized reflected light from the exoplanets \citep[e.g.,][]{sengupta08}. Also, in the case of planets with a very thin atmosphere, the single scattering approximation provides a rough estimate of the observable reflected flux and its state of polarization. However, for a more accurate calculation of reflected flux from a planet and its state of polarization, we need to consider multiple scattering of the internal radiations at the expense of computational time and resources. Figure~\ref{fig:atmi-single}-\ref{fig:atmii-multi} show that the albedo (\alb) increases and $P_{disk}$ decreases by a factor when we consider multiple scattering of the internal radiations for both cloud-free and cloudy atmospheres. Figure~\ref{fig:omega} suggests that the single scattering albedo $\omega$ is a determining factor for $P_{disk}$. The albedo ($F/F_0$) increases with increasing $\omega$ as the amount of light scattered towards the direction of the observer increases. However, the depolarizing effect of multiple scattering of the internal radiations also increases with increasing $\omega$, thereby causing a net decrease in the value of $P_{disk}$. Consequently, in the presence of the atomic absorbers, such as Na and K, the drop in $\omega$ causes the albedo (\alb) to decrease and $P_{disk}$ to increase, as suggested by Figure~\ref{fig:NaK-atmi} and Figure~\ref{fig:hd189733b-NaK}. The effect of $\omega$ shown in Figure~\ref{fig:omega} is in accord with the results of \cite{madhusudhan12}. Also, a comparison between Figure~\ref{fig:atmi-multi} and Figure~\ref{fig:atmii-multi} shows that the peak values of both the disk-integrated flux and polarization phase curves decrease when clouds are present in the atmosphere. The same effect of clouds is also seen in the disk-resolved intensity (scalar) and polarization presented in Figure~\ref{fig:atmi-multi-iscat}-\ref{fig:atmii-multi-iscat}. However, this is not a generic effect of clouds. Analyzing models developed for the hot Jupiter HD 189733 b presented in Figure~\ref{fig:hd189733b}-\ref{fig:hd189733bJ} allows us to study the effect of clouds and other factors in detail as discussed in the following subsections.

\subsection{Analysis of Models Developed for HD 189733 b}

We present models for the hot Jupiter HD 189733 b in order to provide an estimate of the precision required for the detection of this polarization. This planet has been repeatedly targeted for the observational detection and theoretical modeling of the polarization in the optical region \citep[e.g.,][]{bailey18, bott16, wiktorowicz15a, berdyugina11, lucas09, sengupta08, berdyugina08}. We have also made a detailed study of the effects of clouds and the atomic absorbers while developing models for this planet. Figure~\ref{fig:hd189733b}-\ref{fig:hd189733bJ} show the phase curves of \alb, \pdisk, as well as \cont, and \pobs. The quantities \cont, and \pobs\ are more pertinent in the context of observation, as the light from the host star contaminates the reflected light from the planet. We have shown the other quantities to present the dependence of \cont, and \pobs\ on \alb, and \pdisk.

Equation~\ref{eq:pobs-pdisk-albedo-contrast} suggests that \cont\ is directly proportional to \alb, whereas, \pobs\ depends on both \alb\ and \pdisk. \pdisk\, in turn, depends predominantly on $\omega$. The presence of clouds increases the value of $\omega$ at the cloud layers i.e. the layers bound by the cloud-base and the cloud-deck and moreover, $\omega$ increases with an increase in the mean diameter ($d_0$) of the cloud particles. As a result, we find that \pdisk\ decreases with an increase in $d_0$ (see the middle panels of Figure~\ref{fig:hd189733b}-\ref{fig:hd189733bJ}) in all the cases. On the other hand, the albedo (\alb) and hence, \pobs\ depends on both $\omega$ and the optical depth ($d\tau$) of the layers. An increase in $\omega$ causes the albedo (\alb) to increase for a fixed optical depth. On the other hand, if the optical depths of the layers increase (e.g., due to the presence of clouds) causing little to no change in $\omega$, the albedo (\alb) decreases.

\subsubsection{Effects of Clouds at Visible Bands}

In the visible bands, without the presence of the atomic absorbers and clouds, the values of $\omega$ at different layers is found to be extremely high ($>$ 0.99 between TOA and 1 bar pressure level), resulting in a high peak value ($\sim$0.48) of the albedo (\alb) phase curve. Figure~\ref{fig:hd189733b} shows that the maximum observable flux contrast (\cont) and polarization (\pobs) are $\sim$142 ppm and $\sim$22 ppm respectively at 0.55 $\mu$m wavelength. This is in accord with that calculated by \cite{bailey18} and consistent with the observations of \cite{wiktorowicz15} and \cite{bott16}. Further addition of clouds causes only a slight increase in $\omega$. 
For example, $\omega$ at the cloud base level (i.e. the pressure level of $\sim$7500 Pa) changes from 0.9987 (cloud-free) to $\sim$1 (when $d_0=1~\mu$m). At the same time, increasing $d_0$ of the cloud particles also increases the optical depth ($d\tau$) at the cloud layers. The effect of increasing $\omega$ slightly dominates over the effect of increasing $d\tau$ for an increase in $d_0$ up to $d_0\approx0.3~\mu$m. This causes a slight increase in the albedo (\alb) at $d_0=0.3~\mu$m compared to the cloud-free condition (see the top panel of Figure~\ref{fig:hd189733b}). Beyond this value of $d_0$, $d\tau$ of the cloud layers increase substantially causing the albedo to decline for further increase in $d_0$.
However, \pdisk\ monotonically decreases with the increase in $d_0$ as $\omega$ at the cloud layers increase. Hence, due to the combined effect of \alb\ and \pdisk, the peak of the phase curve of \pobs\ is found to be maximum for the cloud-free atmosphere.

\subsubsection{Effects of Atomic Absorbers at Visible Bands}

When we add Na and K in the atmospheric model at 0.55 $\mu$m wavelength, $\omega$ at the same pressure level (i.e., $\sim$7500 Pa) drops to a value as low as $\sim$0.084 in the cloud-free condition. As a result, the peak values of the phase curves of \cont, and \pobs, for a cloud-free atmosphere, are found to be as low as 3 ppm and 0.9 ppm respectively, as shown in Figure~ \ref{fig:hd189733b-NaK}. The presence of clouds increases the values of $\omega$ at the cloud layers significantly (for example, $\omega$ reaches up to as high as $\sim$0.9988 at the cloud base level for $d_0=1~\mu$m). The effect of this steep rise in $\omega$ dominates over the effect of the increase in $d\tau$ at the cloud layers for $d_0\lesssim$ 0.5 $\mu$m, which causes the peak of the phase curve of \alb\ to be maximum at $d_0\approx0.5~\mu$m. On the other hand, the peak of the phase curve of \pobs\ is found to be maximum at $d_0\approx0.3~\mu$m. The corresponding maximum observable values for \cont, and \pobs\ are found to be $\sim$73 ppm (for $d_0\approx0.5 \mu$m) and $\sim$8.25 ppm (for $d_0\approx0.3 \mu$m) respectively.

\subsubsection{Models at NIR Wavelength Region}

Figure~\ref{fig:hd189733bJ} shows that the peak values of the phase curves of \cont, and \pobs\ drop to 10 ppm and 1 ppm respectively in the absence of clouds at $\lambda=1.25~\mu$m. We do not include Na and K in our calculation in this case as the effect of Na and K is absent at this wavelength. As the Rayleigh scattering cross-section gradually falls towards longer wavelengths, $\omega$ and hence, the albedo (\alb) decreases significantly in the longer wavelengths (see Figure~\ref{fig:benchmark-flux}), thereby resulting in these low values of \cont, and \pobs\ in the absence of clouds. Again, the presence of clouds increases the values of $\omega$ at the cloud layers significantly. As a consequence, we find the maximum observable values of \cont, and \pobs\ to be $\sim$83 ppm at $d_0\approx0.7~\mu$m and $\sim$9.25 ppm at $d_0\approx0.5~\mu$m respectively.

This study illustrates that in the low albedo regime (e.g., in the longer wavelengths, or in the visible wavelengths but in the presence of the atomic absorbers, etc.), the detectable polarization can be lower than what has been previously predicted for the planet HD 189733 b. In such a case, future missions have to be more precise than the present capabilities. However, the presence of clouds of a certain mean size can be favorable for the observation of the polarization phase curves for atmospheres with low molecular (Rayleigh) single scattering albedo. Moreover, we demonstrate that we can make a detailed study of the cloud structures and the atomic absorbers, if present in the atmosphere, by comparing the observation of the polarization phase curves in the longer wavelength bands (I or J) with the observation conducted in the visible bands (B, V, or R).  \\
\section{Conclusion} \label{sec:con}

Using the Fortran implementation of our widely used numerical code that solves
the vector radiative transfer equations for calculating stellar and planetary
spectra and the amount of scattering polarization, we have recently  developed
a Python-based numerical package that self-consistently models the atmospheres of
exoplanets and estimates the phase-dependent emergent planetary flux as well
as the phase-dependent polarization profile for any wavelength and for a wide range of
physical parameters of the planet and the host star. In this paper, we have presented the phase-dependent planet-to-star flux ratio  and the 
disk-integrated linear polarization for a spherical reflecting planet with and without the contamination 
from the unpolarized starlight. We have demonstrated the role of multiple scattering in de-polarizing the
reflected planetary light. We have also presented a few benchmark tests for our
numerical package in order to demonstrate the validity of our results. 

 If the star-light can be blocked, the disk-integrated linear polarization due to scattering is quite high in the visible wavelength region.
However, if the unpolarized starlight cannot be separated, the detectable polarization becomes dependent on the planetary albedo. Thus we find that the 
inclusion of the atomic absorbers such as Na and K although enhances the amount of scattering polarization of the disk ($P_{disk}$) in the near optical region, the observable polarization $P_{obs}$ drops substantially because the albedo of a cloud-free planetary atmosphere decreases
in the presence of such absorbers.  On the other hand, the presence of clouds reduces $P_{disk}$, but their effect on the albedo and hence, on $P_{obs}$ depends on the size of the cloud particles, the absorbers present in the atmosphere, and the wavelength band of observation. 

In the visible bands, when Na and K are not present in the atmosphere, the observable polarization is found to be maximum when we consider a cloud-free atmosphere. On the contrary, the presence of clouds of a certain mean grain-size can cause the maximum amount of observable polarization, if we observe the reflected light in the visible bands from an atmosphere containing the atomic absorbers or observe the reflected light in the far optical or in the near infrared bands, where the effect of Rayleigh scattering is extremely low. However, the maximum detectable polarization in the latter case is expected to be a few factors less than the polarization detectable in the visible bands from a cloud-free atmosphere containing an insignificant amount of Na or K.

We have applied our models to the well-studied exoplanets HD 189733 b and predicted the detectable amount of polarization in the
visible as well as in the near infrared wavelength region, considering both cloud-free and cloudy atmospheres. A key conclusion that can be drawn from our entire study is that the observation of the polarized reflected light in the longer wavelength bands (I or J) in synergy with the observation in the visible bands (B, V, or R) by the upcoming missions can excavate valuable information about the atmospheres of the exoplanets.     

We are thankful to the reviewer for having a critical reading of the manuscript and providing useful suggestions. Some of the computational results reported in this work were performed on the high-performance computing facility (Delphinus) of Indian Institute of Astrophysics, Bangalore. We are thankful to the computer division of Indian Institute of Astrophysics for the help and cooperation extended for the present project.

\end{document}